\documentclass[nofootinbib,longbibliography]{paper}
\usepackage{CJKutf8}
\usepackage[greek, hindi, english]{babel}
\usepackage[autostyle, english = american ]{csquotes}
\usepackage[CJKbookmarks, pdftex, bookmarksnumbered, bookmarksopen, colorlinks, citecolor=blue, linkcolor=blue, urlcolor=blue]{hyperref}
\usepackage{amsmath,amssymb}
\usepackage{graphicx}
\usepackage{enumitem}
\usepackage{lipsum, babel}
\usepackage{geometry}
\usepackage{epigraph}
\usepackage{marvosym}
\setlist[enumerate]{topsep=6pt,parsep=-.5mm, leftmargin=7mm,}
\setlist[itemize]{topsep=0pt,parsep=0mm, leftmargin=5mm,}
\setlist[description]{topsep=3pt,parsep=0mm, leftmargin=5mm,}

\def\be{\begin{equation}}
\def\ee{\end{equation}}

\newcommand{\pausa}{\vskip3mm\centerline{\Aquarius }\vskip3mm}

\begin{document}

\title{\LARGE Princeton seminars on physics and philosophy}

\author{Carlo Rovelli}
%\affiliation{Aix-Marseille University, Universit\"e de Toulon, CPT-CNRS, F-13288 Marseille, France.}
%\affiliation{Perimeter Institute, 31 Caroline Street N, Waterloo ON, N2L2Y5, Canada} 
%\affiliation{Santa Fe Institute, 1399 Hyde Park Road Santa Fe, New Mexico 87501, USA} 
%\affiliation{Department of Philosophy and the Rotman Institute of Philosophy, 1151 Richmond St.~N London  N6A5B7, Canada}

\maketitle

\tableofcontents

\section*{Introduction}

\epigraph
{\em We are met as cultivators of mathematics and physics. In our daily work we are led up to questions the same in kind with those of metaphysics; and we approach them, not trusting to the native penetrating power of our own minds, but trained by a long-continued adjustment of our modes of thought to the facts of external nature.}{James Clerk Maxwell{, Address to the Mathematical and Physical Sections of the British Association, 1870}}

\noindent These are lecture notes prepared for a series of seminars I am invited to give at Princeton's Philosophy Department in November 2024.  
As a humble mechanics, I am honored by this invitation. My curiosity for philosophy traces back to early years, but has been nourished by the science I do, quantum gravity.   As Aristotle puts it in the {\em Protrepticus}, ``more in need of philosophy are the sciences where perplexities are greater", and definitely quantum gravity is one of these.  These perplexities, in turn, offer ---it seems to me--- interesting  philosophical suggestions. 
 
I present here a certain number of reflections developed around issues I had to face in my science, and outline the results of a number of papers that might be of interest for a philosopher.  It seems to me that the  scientific revolution of the XX century ---quantum physics and relativistic gravity--- invites us to take conceptual and  methodological steps that have philosophical relevance, as the Copernican-Newtonian revolution did. I am a believer in the fertility of the dialog between physics and philosophy\footnote{CR:  Physics Needs Philosophy. Philosophy Needs Physics, Foundations of Physics, 48 (2018) 481-491.}, and I have had the good fortune of finding fabulous philosophers interested in this dialog and kind enough to listen to my annoying questioning.\footnote{In Pittsburgh John Earman, with whom we planned a joint course on Space and Time, and John Norton;  in Canada Wayne Myrvold, Chris Smeenk and Bill Harper; and then Bas van Fraassen, Jenann Ismael, David Albert, Daniel Dennett, David Wallace, Barry Loewer, Michel Bitbol, Jay Garfield, Mauro Dorato, Enzo Fano, Simon Saunders, Jeremy Butterfield, Erik Curiel, Nick Huggett  and many others. My deep gratitude to them all.}   I offer these views here, humbly,  as a contribution to this dialog. 
  
\pausa

Some ideas recur in these seminars. First, concepts evolve, sometimes radically; to remain anchored to `essential' properties of notions such as time, space, cause, observer, state, system, object, substance, property, fact, perception, observation, consciousness, and many others, is an obstacle to learning.   Second, quantum phenomena and relativistic gravitation suggest a view of the world as a network of relations, rather than being composed of entities with individual properties. Knowledge itself is better understood as part of this same network of relations.  This invites us to a radical perspectival stance: questions about fundamental ontology, or  such as ``does this fact obtain?", may have no absolute sense. Finally, the empirical discoveries of the last century suggest that the best way to think about reality may be to temper foundational ambitions. 

I start with a brief and mostly conceptual account of my central interest in physics, quantum gravity (Seminar \ref{qs}). This is the most `science' of my presentations.  I argue that \emph{space and time are approximate notions} in a quantum dynamics which in general knows neither space nor time. 

Seminar \ref{r} is devoted to quantum physics. I argue that quantum physics can be understood \emph{relationally}, with no need of hidden variables, universal (`many worlds') quantum state, agents, or physical collapse.  But this possibility involves some radical conceptual shifts. 

Seminar \ref{t} is devoted to time, and is organized into three parts.  In the first, I argue that \emph{temporality is a complex phenomenology} composed of distinct {\em layers} that emerge within nested approximations, some of which even involve aspects of our peculiar mental processes. I explain in which sense time is absent in quantum gravity, and why neither Presentism nor the Block Universe metaphor give us good accounts of temporality.  In the second part, I discuss the intuition that time \emph{`flows'} in one direction. I show that this intuition captures the thermodynamic gradient, which is contingent and approximate. Specifically, I  argue that \emph{causation}, meaningless in basic physics, is rooted in this thermodynamic gradient via two complementary paths: the role of dissipation in the temporal orientation of  interventionist thinking, and the deliberating perspective of a subject, which in turn is oriented by the dissipative aspects of its own (physical) mental processes.  The third part discusses what we mean when we say that \emph{the future is `open'}. The openness is traced to the perspective of the deliberating agent. I also briefly comment on the \emph{thermal time hypothesis}, as a possible tool for relating the first part with the second and third. I also comment on the status of the speculative hypothesis of a \emph{perspectival understanding of the arrow of time}. 

Perspective takes center stage in Seminar \ref{subject}, where I present some comments on the role of the \emph{the subject} in understanding the natural world. I illustrate some clues that physics offers for bridging physical and mental notions such as \emph{information} and \emph{meaning}.  I argue that we better not only consider that the subject of knowledge is part of nature, but also that \emph{knowledge is itself a physical phenomenon} (captured in descriptive rather than normative terms): a peculiar form of correlation. I argue that the subjective perspective (and its associated phenomena) is not a mystery: it should universally underpin a mature physicalism. I also argue that perspectival knowledge is not second-rate knowledge, because all knowledge is naturally realized, hence imbedded in the world.  The ambition of non-perspectival knowledge introduces the dualism that generates puzzles and aporias. 

Seminar \ref{pp} draws some methodological and general observations. I argue that continuity is under-appreciated in physics and this is currently misleading research.  I emphasize the role of acquired  knowledge, and of analogical and metaphorical thinking, in the development of concepts.  I summarize the general views on conceptual evolution, relational worldview, and anti-foundationalism, to which it seems to me today's physics is inviting us. 

\section{Quantum spacetime} \label{qs}

\epigraph
{\em  If anyone believes that certain concepts are absolutely the correct ones, and that having different ones would mean not realizing something that we realize, then let him imagine certain very general facts of nature to be different from what we are used to, and the formation of concepts different from the usual ones will become intelligible to him.}{Ludwig Wittgenstein, Philosophical  Investigations, II, 12}

\noindent  This first Seminar has more physics and less philosophy  than the others. Yet, I  think it is also the most philosophically challenging, because quantum gravity requires the kind of conceptual shifts  that I will argue later to be a core component of science.  The reason is that the way space and time are conceptualized in the theory is a departure from the way they are in Newtonian mechanics or special relativistic classical or quantum field theory.   

I also argue, however, that the departure is perhaps not so radical after all, since to some extent it is just falling back to Cartesian--Aristotelian notions. 

\pausa 

Quantum gravity is a concrete problem. We see matter falling towards the many black holes that --we now know-- dot the sky: where does it go?  We expect these holes to ``evaporate": what happens at the end of the  evaporation?  We have a rather convincing reconstruction of what happened on large scales during the last 13 and so billion years: what happened earlier?   Quantum gravity is the problem of answering these questions by finding a mathematical and conceptual structure accounting for these phenomena.  

The problem has philosophical relevance because gravity determines what we call the `geometry' of space and time. (This is the core content of Einstein's General Relativity, or GR, now become one of the best empirically confirmed scientific theories ever, as recognized by a recent stream of Nobels.)  Quantum properties of gravity are therefore quantum properties of space and time.  This requires us to reconsider the way we think about space and time.  The right question, as I see it, is not the ill-defined questions: What is space? What is time?  Rather, it is, less pretentiously: what is the clearest way, available to us today, to think about space and time, in light of what we have learned from the empirical sciences?

\pausa

The conceptual structure of GR is clean and clear (unlike quantum theory), but subtle and often misrepresented. 

GR describes reality as an ensemble of {\em fields}.  Fields are extended entities.  In GR, they are not sitting ``in space" or ``on a space", or evolving ``in time".  Rather, they {\em define} spatial and temporal extension themselves,  much like Descartes's ``res extensa" was not ``in space", but rather defined space, thanks to its property of being `extended'.\footnote{The way this is realized in the mathematics is via the invariance of the field equations under general coordinate transformations.   This implies that the GR coordinates ($\vec x, t$) must necessarily be interpreted in a profoundly different manner than the familiar Euclidean coordinates ($\vec X,T$) describing space and time in pre-general-relativistic theories.  While the second are assumed to capture the geometry of space and time, the former do not.  The fields are defined on a manifold coordinatized by ($\vec x, t$), but the invariance identifies any two configurations that are the same up to the way they sit on the manifold: this invariance ``washes away"  the manifold.  What remains physically relevant is not where things (such as an excitation of a field) are with respect to the manifold, but only where things (including the gravitational field) are located with respect to one another.}   

Among the fields, there is the {\em gravitational} field, which is just one among the interacting dynamical entities of the theory. The processes this field  undergoes are christened ``spacetimes".  The reason for this peculiar naming is that (in the light of GR) the Euclidean space of Newton's mechanics and the Minkowski spacetime of special relativity are special configurations that the gravitational field may assume.\footnote{Respectively within and outside the non-relativistic limit.}   

To get clarity about this subtle conceptual shift, and the role it plays in quantum gravity, we must distinguish two \emph{distinct} commonly-employed meanings  of the word ``space"  (I postpone the discussion of the general relativistic notion of \emph{time} to the next section):

\begin{enumerate}[label=(\roman*)]
\item By ``space" we can refer to the set of relations that concern the location of things, namely the question ``where?"  The common answer to ``where is $A$?" is to name  things in the vicinity, adjacent, or surrounding $A$.   Where are you? I am in the Princeton Philosophy Department.  This defines a location, and space is the ensemble of all possible locations.  Since location of an object, in this sense, is defined by its surrounding objects, location is a {\em relational} notion. That is, it is a property determined together by both the located object and the objects with respect to which the object is located.  This is the old Aristotelian definition of space\footnote{The location of an object is the innermost  boundary of the surroundings (\textgreek{Φυσικά} IV.5).}, as well as the one utilized by Descartes\footnote{Location is the neighborhood (\emph{vicinia}) of the bodies immediately contiguous (Principles of Philosophy II, 25).}, and is also the most commonly employed in everyday life.  For lack of a better term, this can be denoted {\em relative space}, or \emph{relational space}, as it is defined by the {\em relation} of adjacency between things. 
\item After Newton, ``space"  has acquired also a second meaning, today very familiar to physicists and philosophers alike.   It designates an element of reality, which Newton assumed to have an existence independent from the objects ``in it": a self-standing entity with geometrical properties, with respect to which the locations of all other entities are preferentially defined. To write the Newtonian equations describing the motion (hence the changing {\em location}) of a particle, we do not need other entities to be present in the universe, besides space, time, and the particle itself.  For lack of a better term, this can be denoted,  ``physical space", or ``Newtonian" space, as it is really a discovery by Newton.
\end{enumerate}

To understand GR properly, it is important to keep this distinction clear\footnote{It is close the famous one at the beginning of Newton's Principia. The first is called by Newton relative, apparent, common; the second absolute, true, mathematical, terms carrying a heavy value judgement.}.  In terms of this distinction, GR is the discovery that the entity  that Newton hypothesized, namely \emph{Newtonian space}, is indeed an entity independent from the rest of matter (as Newton correctly understood\footnote{It determines which state of rotation has physical effects, as in his celebrated bucket experiment, and how a ``free" particle moves.}), but it is not a fixed structure (as Newtonian physics mistakenly assumes): rather, it is a dynamical field, in many ways similar to the electromagnetic and the other fields. It has its own dynamics and obeys field equation like all dynamical physical fields.   

On the other hand, the  \emph{relative} notion of space, the one of our everyday life, is {\em not} affected by GR: there still is a structure of adjacencies, we still locate objects with respect to one another. One of them is the gravitational field itself. This is GR's relationalism: location is fully relational.\footnote{Once this reconceptualization of space is acquired, the debate between substantivalism and relationalism seems to me to reduce to terminological choices (CR, Quantum Gravity, II.4.2, CUP 2014). By denoting a gauge equivalence class of metrics as ``spacetime", we fulfill the substantivalist's expectation of spacetime as an entity (H Halvorson and and JB Manchak, Closing the Hole Argument, 2021).  Alternatively,  denoting it as ``gravitational field", and realizing that {\em unlike special relativistic physics}, general relativistic physics implies that dynamical entities are only localized with respect to one another, we reveal the peculiar relational nature of localization in the theory. This nature is emphasized by the ``hole argument", which (correctly) motivated Einstein to conclude that there is a sense in which physical points do not have an identity beside that given them by the dynamical fields.(A Einstein,  Relativity and the Problem of Space, Appendix V, added (in late editions) to Relativity, the Special and General Theory, Routlege 1952, pp. 135-157).  (To further muddle matters, the same gauge transformations are called ``diffeomorphisms" or ``isometries" in different papers.)}\\[1cm] 

\pausa

Let's bring the quantum phenomena into this conceptual structure.  Since the gravitational field is a field like any other, there is no plausible reason to imagine it \emph{not} to be affected by quantum phenomena, as all other fields are.\footnote{Here I am relying on methodological assumptions that I discuss in the last Seminar: we have plausible reasons to make educated guess about domains we do not yet have empirical access to, by relying on incomplete past knowledge. More than that: we better take current knowledge seriously.  Here the situation is the following. We have learned from GR that spacetime is a dynamical field. We have learned from quantum mechanics (QM) that dynamical fields have quantum properties.  The colossal empirical success of these two theories motivates us (by induction) to tentatively trust their core discoveries beyond the regimes where they have been tested. These discoveries are like the discoveries that velocity is relative, that the Earth is not the center of the universe and the Sun does not rotate around the Earth, that there are physical fields, that at low velocities and low distances massive bodies attract one another gravitationally, that all animals have common ancestors,  and so on.  These are \emph{established} elements of our knowledge of the world: they are the best guide we have for the unknown. This is why, by the way, I consider speculations like hypotheses that gravity is insensitive to QM, or QM breaks the general covariance of GR, to be of scarce interest.} Einstein himself, only one year after completing GR, pointed out that GR \emph{must} be incomplete, because it disregards quantum phenomena.  There must be a coherent quantum description of the gravitational field, of which GR is the non-quantum (or $\hbar\to 0$) limit.

Quantum phenomena are characterized by three features:  
\begin{enumerate}[label=(\roman*)]
\item physical variables can have {\em discrete} spectrum; 
\item their dynamics can only be predicted {\em probabilistically}; and 
\item  values of variables are {\em contextual}: they depend on the physical context in which the system in question finds itself. More precisely, they are {\em relative} to other systems, namely those the system is interacting with.\footnote{I discuss this contextual and relational aspect of quantum theory in detail in the next Seminar.} 
\end{enumerate} 

Hence we expect the quantum properties of gravity to imply that physical space evolves \emph{probabilistically}, and its properties to be \emph{discrete} and \emph{contextual}.  

 A mathematical\footnote{See for instance CR, Francesca Vidotto, {Covariant Loop Quantum Gravity}. Cambridge University Press 2015.} and conceptual\footnote{See CR, Francesca Vidotto, ``Philosophical Foundations of Loop Quantum Gravity", in Handbook of Quantum Gravity, C Bambi, L Modesto and I Shapiro editors, 2023, Springer. arXiv:2211.06718.} structure to account for these features is provided by Loop Quantum Gravity (LQG).  LQG is a rather well defined theory of quantum gravity.  Whether, and the extent to which, it precisely describes Nature is a question that will be decided by empirical confirmation (this is the tool upon which science's reliability depends). But as a possible coherent unification of the two revolutions in physics of the XX century, GR and QM, I think it is of philosophical interest as it is, because it shows what the two discoveries imply, when taken together.  To see how it works, I illustrate how it does so in a concrete example.   

\pausa

Say we are interested in a specific quantum gravitational process: the formation, evaporation and possible final dissipation of a black hole. Imagine for concreteness that we had the technology to produce small black holes, say a few micrograms in mass, in the lab.\footnote{We do not have ---by far--- the technology to do so: we would need to concentrate this mass in far too small regions. But in principle this is possible.}  We could wait and see what happens.  How would we measure this process and describe it theoretically? 

\begin{figure}[t]
\centerline{\includegraphics[width=8cm]{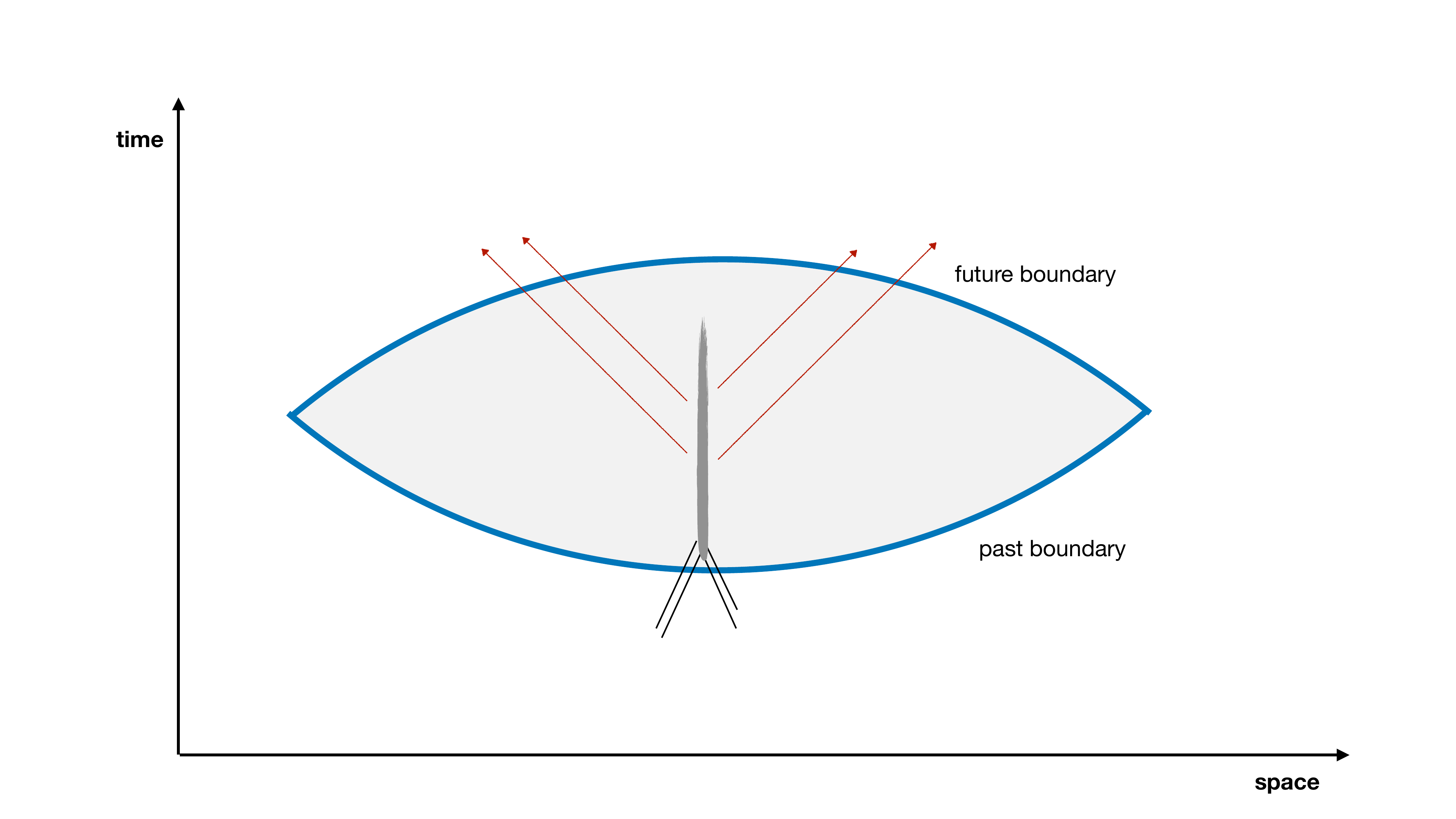}}
\caption{2d scheme of a quantum gravity process: the blue line represents the (3-dimensional) boundary of the (grey) spacetime region forming the process.}
\label{f1}
\end{figure}

The  process happens within a limited spatial region and within a limited lapse of time.  That is, the process happens in a limited region of spacetime.  This region has a boundary. Let us focus on such {\em boundary} of this spacetime region.\footnote{One may object: ``Carlo, why are you using spacetime notions, if you want to get rid of these notions?". It is a good question, of course.  The answer is in Neurath's simile. Please hold on and bear with me until the next Seminar, to get full clarity.}   This can be chosen to be formed by two space-like surfaces, in turn joined at their boundary, as sketched in Figure \ref{f1}.

Imagine we have detectors situated along the blue line of the figure. They measure all relevant fields before and after the process. One of the fields of course is the gravitational field. Let us here disregard  the other fields for simplicity of presentation.  

\begin{figure}[h]
\centerline{\includegraphics[width=5cm]{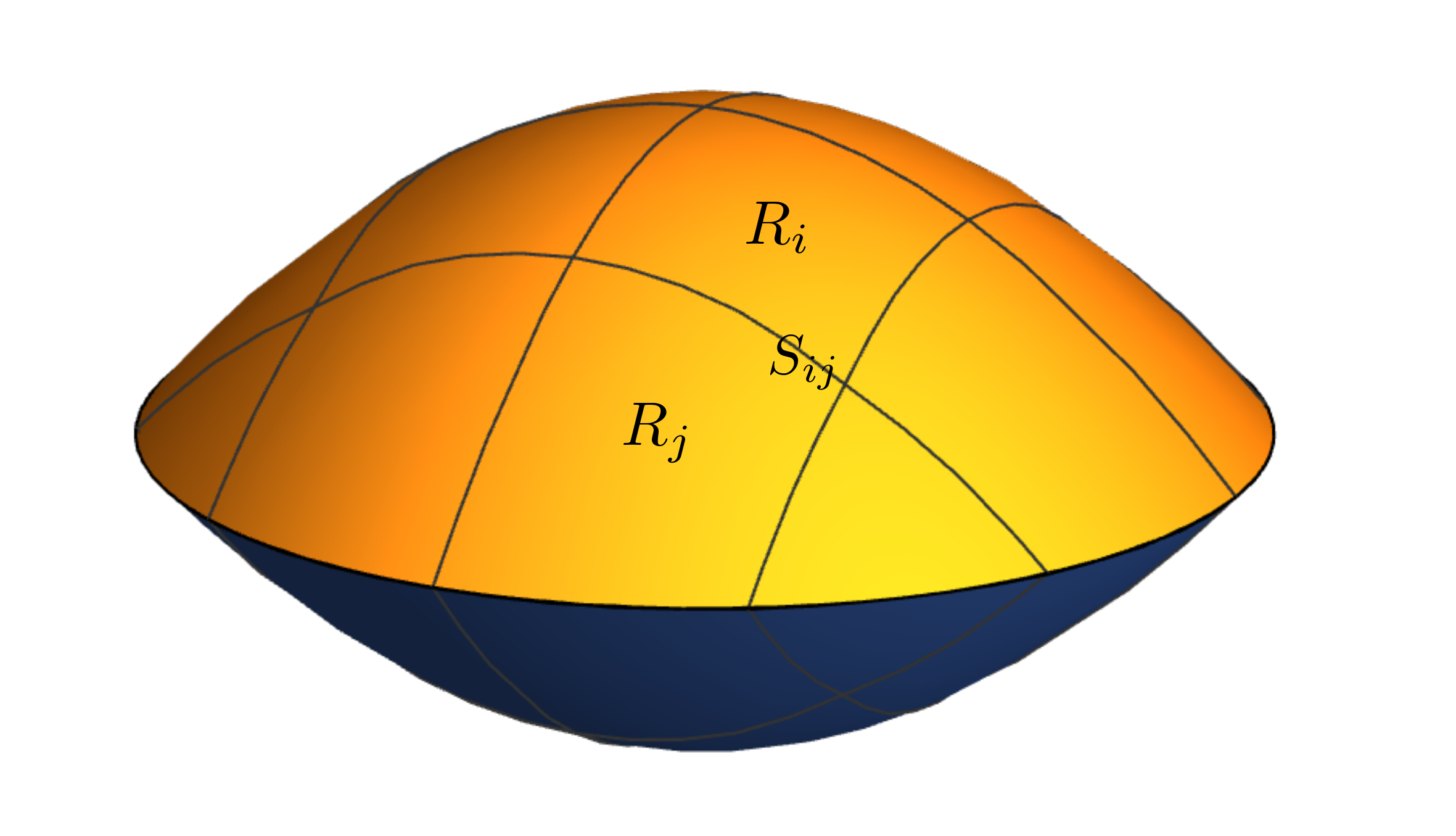}}
\caption{3d rendering of the boundaries enclosing the process, which corresponds to the blue line of  Figure \ref{f1}, and its cellular decomposition. }
\label{f2}
\end{figure}

Since measuring the gravitational field is synonymous to measuring the geometry, we are also measuring the geometry on the surface bounding the process.  Geometry is continuous, hence we cannot measure it exactly, as this would require gathering an infinite amount of information.  We can rather imagine, for concreteness, approximating the geometry using a truncation. This can be done partitioning the past  and future boundary surfaces in regions $R_i$, and measuring the volumes $V_i\equiv V(R_i)$ of these regions and the areas $A_{ij}\equiv A(S_{ij})$ of the surfaces $S_{ij}$ separating the regions $R_i$ and $R_j$. See Figure \ref{f2}.  If we schematize each region as a point, and we connect any two adjacent regions with a link between the two respective points, we obtain a graph, that I denote $\Gamma$.

The quantities $(V_i,A_{ij}, \Gamma)$ provide a (truncated) amount of information about the beginning of the process happening in the grey region and the analogous quantities $(V'_k,A'_{kl}, \Gamma')$ are a (truncated) amount of information about the end of the process in the grey region. 

A quantum theory of gravity must  then give two specifiations about reality:
\begin{enumerate}[label=(\roman*)]
\item A list of the possibly discrete values that the quantities $V_i$ and $A_{ij}$ can take. 
\item The probabilities for the transition $(V_i,A_{ij},\Gamma)\longrightarrow (V'_k,A'_{kl},\Gamma')$. 
\end{enumerate}
Loop quantum gravity defines the ingredients for computing these quantities.  

\pausa

The technical way this is realized is the following. The theory is defined by a triplet $(\cal A, \cal H,  W)$, where $\cal A$ is an algebra formed by quantities with a physical interpretation (in particular, $\cal A$ contains the Area and Volume variables associated to a truncated measure of the geometry of a surface as described above),  $\cal H$ is a Hilbert space of states on this algebra, and  $W$ is the operator that defines the transition probabilities.\footnote{In extreme synthesis, the Hilbert space can be taken as a suitable $n\to
\infty$ limit of the spaces $H_n=L^2[SU(2)^{\frac{n(n-1)}2}/SU(2)^n]$ associated to the complete graph with $n$ nodes, and the dynamics is determined by the spinfoam vertex amplitude 
 $ A_v(\psi) = P_{SL(2,C)} Y_\gamma \psi(1\!\!1)$, where $P_{SL(2,C)}$
 is the projector on $SL(2,C)$ singlets and $Y_\gamma$ is the map from $SU(2)$ to $SL(2,C)$ representations that reads 
$|j;m\rangle \mapsto   |\gamma j,j;j,m\rangle$ in the canonical basis. For a complete mathematical definition, see for instance CR and F Vidotto, Covariant Quantum Gravity (CUP 2014).  }

The values that the quantities $A_{ij}$ and $V_i$ can take (see (i) above) are the elements of the spectrum of the corresponding algebra element. The transition probabilities (see (ii) above) are the modulus square of a transition amplitude given by the matrix element:   
\be
W(V_i,A_{ij}, \Gamma;V'_k,A'_{kl}, \Gamma')=\langle v'_k,j'_{kl},\Gamma|W|v'_i,j'_{ij}\ \Gamma'\rangle
\ee
where the quantities $|v_i,j_{ik},\Gamma\rangle$ are eigenstates of quantum operators with quantum numbers $v_i,j_{ik}$ corresponding to the eigenvalues $V_i,A_{ij}$. This defines the theory. 

\pausa

The spectral problem for the Area and Volume variables has been solved in LQG: the eigenvalues of these quantities have been computed.   Their spectrum is discrete. This is a main result of LQG.  Eigenvalues of the Area are given by
\be
A_{j_n}=8\pi G\hbar\gamma \sum_n \sqrt{j_n(j_n+1)},
\ee
where the quantum numbers $j_n=0,1/2,1,3/2,2,...$ are half integer for any $n$, the constants $G$ and $\hbar$ are the Newton and Planck constants, and $\gamma$ is a dimensionless parameter that enters in the definition of the theory.  This is the ``quantization" of the geometry, in the sense of ``discretization". 

Therefore the theory predicts that {\em any} area we measure would come out to have one of the discrete values $A_{j_n}$, for some set of half-integers $j_n$.  For instance, a scattering measurement measures a cross section.   Cross sections are areas.  The theory predicts that any  cross section measured with enough precision would turn out to be one these numbers.  This is analogous, for instance, to the prediction that any angular momentum is discrete, in non-relativistic quantum theory.   Or the energy of any oscillator being similarly quantized. 

This result is central in LQG, and has direct physical consequences. For instance, it indicates the possible existence of quasi-stable objects enclosed by a horizon of the minimal non vanishing area. The mass of these objects is of the order of a fraction of a milligram (the ``Plank mass").\footnote{\href{https://arxiv.org/abs/2309.08238}{M Christodoulou et al, Detecting Gravitationally Interacting Dark Matter with Quantum Interference}, 2023.}

The quantum states that diagonalize areas and volumes are called ``spin network" states.\footnote{The term network refers to the graph $\Gamma$, the term spin refers to the quantum numbers $j_{ik}$; these are labels of representations of the rotation group, usually called `spins'.}  The operator $W$ defining the transition amplitudes\footnote{A generalized projector on the Hilbert space.} is defined order by order in a truncation scheme. At every order, its matrix elements are constructed by integrals over matrix elements of unitary representations of the Lorentz group.  Here these technical details are not relevant.  As an example, the amplitudes have been used to estimate the probability for a black hole to quantum tunnel into a white hole. (The calculation indicates that this probability approaches unity at the end of the Hawking evaporation.\footnote{\href{https://arxiv.org/abs/1801.03027}{D'Ambrosio, Christodoulou, Characteristic Time Scales for the Geometry Transition of a Black Hole to a White Hole from Spinfoams}, 2018, 	arXiv:1801.03027.})  

\pausa

The above is the outline of a complete theory of quantum gravity.  It is tentative because we still lack empirical support.  From the philosophical perspective it is a good example of what GR and QM can imply, once taken together. 

The resulting structure of quantum gravity is  similar to that of elementary quantum theory in many respects: the theory is defined by {\em variables} captured by the algebra, {\em states} coded in the Hilbert space, and a {\em dynamics} given by $W$, like any other quantum theory.   

But several peculiarities are conceptually relevant:

\begin{enumerate}[label=(\roman*)]

\item There is {\em no explicit temporal variable} in the theory.  

\item The {\em Heisenberg cut} (namely the artificial and largely arbitrary boundary that separates the quantum process considered and the rest of the universe, or the ``measuring apparatus") {\em is identified with the boundary of a spacetime region}. 

\item There is no spacetime geometry associated to the region enclosed inside the boundary.  Only the transition amplitudes have physical meaning.  {\em The spacetime region enclosed within the Heisenberg cut is the quantum process itself}. 

\item Generic boundary states can be in generic quantum superpositions of eigenstates.  This happens in particular when the measured quantity is a variable conjugate to the geometry of the boundary.  That is, {\em the boundary geometry can be in ``quantum superposition"}, like the position of a quantum particle.  

\item The only geometric structure in the theory is given by the values of areas and volumes and their adjacency relations (at a given truncation).  {\em No other spatial relation} is evoked.

\item Quantum gravity is the quantum theory of a specific entity: the gravitational field. Quantum gravity does not need to postulate other entities. 

\item The theory describes how a certain spacetime process manifest itself {\em to something external}. 

\end{enumerate}

\vspace{3mm}

The various conceptual issues raised by these features will be addressed in the course of these Seminars. I start here with some simple notes on the notions of space and geometry. 

\begin{description}
\item[Boundary geometry and relational space.]  Geometry is not determined by an external structure as in Newton theory, special relativity, non-relativistic quantum theory, or quantum field theory. Rather, it is  part of the kinematics, like in classical  GR.   \emph{Variables are not located in spacetime}: they define location themselves.  Localization is defined relatively to other dynamical objects in the theory. This is the Aristotelian-Cartesian notion of space: things are located with respect to one another, not ``in space".   This is radical, but, as previously discussed, it is already so in classical GR. 
 
 \item[Quanta of space.] The ``things" we are talking about, if there is no other field beside the gravitational field, are the individual regions represented by the points of the graph, and their geometries.  These regions have quantized volumes. They are called ``quanta of space", following the terminology for the ``quanta of light" that Einstein introduced in 1905.\footnote{Careful: quantum discreteness is not in the discreteness of the nodes: it is the discrete values of their volume and area. The first is just a truncation of degrees of freedom, like taking a finite number of Fourier components of the electromagnetic field (no $\hbar$ here).  The second is quantum discreteness, like the quantization of the energy of each Fourier component of the field ($E=h \nu$). Mixing up the two is a common source of confusion.}  Prima facie, these can be taken as the ``entities" the theory describes. They have analogies with Einstein's quanta of light, namely the photons, or the particles of quantum field theory, which are discrete quantum excitations of a quantum field as well.  An important difference is that while the photons or the particles of quantum field theory are located in space, the quanta of space are themselves space (in the Newtonian sense of ``space")\footnote{This is reflected in the fact that the quantum number of a photon includes its position --or its Fourier transform: its momentum-- while there is nothing similar for a quantum of space, whose quantum numbers are its area, volume and its adjacency relations with other quanta of space.}.  Newtonian space is re-interpreted as an approximate description of a sea of these quanta.\footnote{LQG can be easily coupled to external fields, like fermions of electromagnetism.  In this case, the particles and the photons simply interact with the quanta of space of LQG, instead of being immersed into a continuous spacetime.} However, such prima facie reading of the theory as a theory of `objects' represented by quanta of space is misleading.  This is so for the same reason for which photons, or the particles of quantum field theory, are not objects: they are modes of interaction of the field with an apparatus, a screen, of something else.   In precisely the same manner, in LQG the quanta of space are only defined at the boundary, where the process in question interacts with the other side of the Heisenberg cut. Quanta are not things.  Furthermore, they depend on the truncation chosen, which in turn depends on the external variables in terms of which we describe the process.  I come back to this subtle and crucial point after discussing quantum mechanics in general in the next Seminar.   

\item[Time.] A striking feature of the theoretical scheme of LQG is the absence of a variable that can be readily identified with time. I postpone the discussion of this point to the third Seminar,  devoted to time.  I only anticipate here that in no way does this feature of the theory imply that ``time is frozen" or that we must think of the universe as a four dimensional present, or any similar nonsense.  The theory is about processes, happenings, events, all notions that we commonly call temporal.   Indeed, the region enclosed in the boundary is to be interpreted as a {\em process}, in the sense in which a soccer match or a nice dinner are processes.  In usual parlance, processes are happenings confined in a region of spacetime.  With respect to the external world, what I have described above is precisely a process in this sense.  Physics is not about things: it is about processes. 

\item[Bulk geometry.] There is no spacetime geometry in the region enclosed within the boundary.  This should not be surprising.  In QM it is always the case that the notion of classical trajectory loses meaning.   An electron, for instance, transitions from one Bohr orbit to another via a quantum leap. Or: it makes no sense to ask which of the two slits a particle has crossed, in a double slit interference quantum demonstration.    In the case of GR, a spacetime, namely a portion of a solution of the Einstein's equations, is a trajectory of a dynamical entity. Like all trajectories, this solution loses its meaning in the quantum regime.   A physical quantum gravitational process cannot be represented in classical spacetime terms.  There is no metric in it.\footnote{In terms of Feynman's  sum-over-histories intuition, we can picture in the mind a quantum gravity process as a ``quantum sum" of all possible bulk geometries interpolating the boundary geometries, each weighted by a complex phase given by its action.  Of course this is not a representation of reality: it is a representation of a calculation.  It is nevertheless a useful support for intuition (first suggested by John Wheeler and later developed by Stephen Hawking).}   As a quantum particle can perform a quantum leap between locations, so space can quantum leap between geometries.   In a precise sense, physical spacetime is locally replaced by quantum gravitational processes. Spacetime {\em is} a quantum gravitational process.   

\item[Observables on spacetime boundaries.]  The boundary of a spacetime region is identified with the Heisenberg cut of the quantum process defining this region.  The observable quantities are defined on this boundary. 

\item[Adjacency=interaction.] Two processes affect one another only if they are adjacent. On the other hand, adjacency --the relation that weaves spatiality-- is only manifested by the possibility to interact. In this sense quantum interactions (and hence the entanglement they generate) appear to be building spatiotemporal adjacency.  This is not sufficient to generate gravity (as sometimes speculated), because gravity is mediated by the gravitational field. 

\end{description}

\pausa 

A much discussed conceptual difficulty in quantum gravity is to identify local ``observables", namely good variables for defining the theory.  These have to respect the gauge invariance of the theory. In GR, this means that they have to be defined relative to the location of something; in QM, they must be defined relative to another system (Bohr's `apparatus', see next Seminar). These two requirements, taken together, lead to the idea of {\em identifying the spacetime boundary of the quantum process with its Heisenberg cut}. Here is where the relationalism at the core of GR meets the relationalism at the core of QM.   The first is the fact that localization is only defined relative to something else.   The second is the fact that the quantum processes of a physical system are only defined relative to something else.   It is the convergence of these two relationalisms that makes quantum gravity coherent, as I understand it.  The first stems from the discussion above about the notion of space in GR (which has, as we shall see in Seminar \ref{t}, a perfect analog for time).   The second is discussed in the next Seminar.\\[1cm]

 \section{Relations} \label{r}

\epigraph{\em 
Emptiness is the elimination of all views. \\
Anyone for whom emptiness is a view\\
Is incorrigible.}{Nāgārjuna, Mūlamadhyamakakārikā, 13.8}

Quantum gravity is a quantum theory.  As such, it inherits the obscurity of standard quantum theory.  We cannot get clarity about the first, without conceptual  clarity about the second.  

There is no consensus about the best way to think about quantum phenomena. Different ways to do so are considered by physicists and philosophers, under names such as Many Worlds, De Broglie-Bohm, physical collapse, and similar.  Several of these ways of conceptualizing quantum phenomena seem viable to me in principle, but I do not judge them equally fruitful, equally promising, or equally suitable for quantum gravity.  The way of thinking about quantum phenomena that seems to me more plausible and fruitful, both in itself and in view of quantum gravity, is the Relational Interpretation.\footnote{\href{https://arxiv.org/pdf/quant-ph/9609002}{CR, Relational quantum mechanics}, International Journal of Theoretical Physics 35 (1996) 1637--1678.  \href{https://www.princeton.edu/~fraassen/abstract/Rovelli_sWorld-FIN.pdf}{Bas van Fraassen, Rovelli's World}, Foundations of Physics 40 (2010), 390--417. \href{https://plato.stanford.edu/entries/qm-relational/}{F Laudisa, CR, Relational Quantum Mechanics}, Stanford Encyclopedia of Philosophy.  \href{https://arxiv.org/abs/2109.09170}{CR Relational Interpretation}, in Oxford Handbook of the History of Interpretations of Quantum Physics (Oxford UP, 2022).}  

The Relational Interpretation does not interpret the current confusion about quantum theory as a sign that what is necessary to render the theory intelligible is a new equation (as in De Broglie-Bohm theory), some not-yet observed phenomena (as in the physical collapse hypotheses), or the assumption of the existence of a vast inaccessible domain of reality (as in the Many Worlds' universal quantum state.)  Rather, it interprets quantum phenomena as an invitation to a radical update of the conceptual framework we use to think about reality. 

I discuss it here because it explains aspects of the theory defined in the previous Seminar, because of its intrinsic interest, and because it is a pillar of the general perspectival view of physics that I understand the XX century revolution to imply.

 \paragraph{Quantum theory as a theory of relations.} \label{q}
The relational interpretation (or RQM, for Relational Quantum Mechanics) is the idea that the variables that can be used to describe a physical system only take value when this system is acting upon another system, and the value taken is only relative to that other system.  More sharply, for a system to be described by a variable taking a value is to affect another system in a certain manner, \emph{and nothing else}. 

This is an idea that breaks away from a venerable conceptual structure that has long underpinned physics in particular and perhaps our worldview in general: the idea that the world can be thought as composed of systems that  at any time have variable properties by themselves, irrespectively of any other system, and these properties can be described by the values taken by variables characteristic of the system in itself.    The break away is double.   First, in general the variables that characterize a system \emph{do not have a value at all times}; that is, there are times at which a variable takes no value.  Second, the value that a variable takes is always \emph{relative to another system}, the system it is interacting with.

This assumption permits resolving a number of apparent paradoxes raised by quantum physics\footnote{For a nice illustration of RQM applies to common apparently paradoxical situation in quantum mechanics, see \href{https://philsci-archive.pitt.edu/23375/}{V. Fano, Application of Relational Quantum Mechanics to Simple Physical Situations}, pittphilsci:23375, 2024.}, and brings  quantum phenomena back into a coherent story  --although a story with a radical conceptual novelty-- without the need of invoking many worlds or hidden variables \`a la De Broglie-Bohm, or interpreting quantum theory as necessarily incomplete.  

Here are two examples of the application of this view of the theory: 

 In the double slit setting, the position of the quantum particle takes a definite value with respect to the screen hit by the  particle, at the time the screen is hit.  But the position at the time at which the particle `crosses the holes' has no value. 
  
In the Wigner's friend setting, a variable $A$ of a quantum system measured by Friend takes a value, say $a$, with respect to Friend, but not with respect to Wigner.  Concretely, this means that any subsequent value taken by variables of the particle and Friend \emph{relative to Wigner} are not influenced by the value taken by $A$ relative to Friend. Specifically, the probabilities of the different values these variables can take are affected by quantum interference between different values of $A$.   In simpler words, $A$ can take a value with respect to Friend and still be ``in a quantum superposition" with respect to Wigner.  This is the central idea of the relational understanding of the quantum phenomena. 

Physics is about processes. Quantum theory accounts for processes by means of transition amplitudes. The entries of the transition amplitudes are values of variables that are always (often implicitly) \emph{labelled} by the system with respect to which they take value. The insight of RQM is that transition amplitudes have physical relevance only when all entries are labelled by the same system.  That is, if $a_F$  (or $b_F$) and $a_W$ (or $b_W$) are values of a variable with respect to $F$ and $W$ respectively, then 
$ W(b_F, a_F) $ (same label) has physical meaning, but  $W(b_W, a_F)$ (different label) does not. 

It is this stipulation that allows interference, namely the failure of the following relation between probabilities (here $P(a|b)=|W(a,b))|^2$):
\be
P(c_W|a_W)=\sum_{b_F} P(c_W|b_F)P(b_F|a_W).
\ee
The terms in the r.h.s.~of this equation are \emph{not} predictions of the theory.  The failure of the above equation is what we call quantum interference, and is due to the fact that the proper quantum relation is of course 
\begin{eqnarray}
P(c_W|a_W)&=&|W(c_W|a_W)|^2=\left|\sum_{b_F} W(c_W|b_F)W(b_F|a_W)\right|^2\nonumber \\
&=&\sum_{b_F} P(c_W|b_F)P(b_F|a_W)+{\rm interference\  terms}.
\end{eqnarray}
The main source of the obscurity of the quantum phenomena is the apparent contradiction between the existence of determined facts in nature (the ``measurement outcomes") and the ubiquitous quantum superposition implied by quantum dynamics.  Attributing values only at interaction and understanding them as labelled resolves the contradiction. 

\pausa

Comments:
\begin{description}

\item[Systems and variables.] The world can be viewed as a collection of physical systems undergoing processes; a physical system is characterized by a family (technically, an algebra) of variables.  Variables take values relative to another system, when the two systems affect one another. The value taken describes the  manner in which one system affects the other. The specifically quantum properties of a variable are twofold: (i) Not all values of the variables are physically realizable; only certain ``quantized" values are\footnote{Mathematically, these are determined by the spectrum of the corresponding algebra element.}, and: (ii) The dynamics fixes only probability relations of the distribution of these possible values in time.\footnote{Mathematically, these are determined by moduli square of transition amplitudes. Transition amplitudes are scalar products between observable eigenstates in a representation of the algebra, in the Heisenberg picture.} In the previous Seminar I have illustrated how this is realized in LQG.

\item[States.]  The notion of ``state" is \emph{not} part of this basic conceptual setting.   A physical system has no state by itself.   A quantum state $\psi$ can still be defined as a mathematical tool coding the information about ways a system $A$ has affected another system $B$.  Defined in this manner, $\psi$ is a state of $A$ \emph{relative to $B$}; it is a calculation device useful for predicting, on the basis of the information it codes, the (probabilistic distribution of the) possible ways $A$ can affect $B$.    This role is analogous to the role played by the Hamilton-Jacobi function $S$ in classical mechanics. In fact, the  Hamilton-Jacobi function can be obtained from the semi-classical approximation of the quantum state ($\psi\sim e^{\frac{i}{\hbar}S}$).    
 
 The realization that the quantum states $\psi$ concretely used  in physics are always {\em relative} states is a major  insight by Hugh Everett.\footnote{H Everett, Relative state formulation of quantum mechanics. Rev. Mod. Phys. 29 (1957) 454–462.}    Everett's insight has also been developed in a direction different from RQM: the direction of what is today called the Many Worlds interpretation.   In this interpretation, Everett's relative states $\psi$ (the ones used in calculations and in the lab) are \emph{assumed} to be components of an absolute (non-relative) universal quantum state $\Psi$, to which a strong ontological commitment is generally declared.  In the Many Worlds view, that is, reality \emph{is} taken to be $\Psi$, within which perspectival values of variables and relative states $\psi$ are associated to systems and branches of $\Psi$.  It seems to me that relative quantum states could be spared any ontological commitment (who would assign ontological weight to the Hamilton-Jacobi function?), the universal quantum state $\Psi$, which plays no role in the actual use of quantum mechanics, can be dropped, and we can content ourselves with the perspectival values of variables.  The only role of the universal $\Psi$ is to satisfy a metaphysical prejudice, which we can drop.  
 
 \item[Matrix mechanics versus wave mechanics.] The emphasis on values of variables rather than quantum states is closer in spirit to first formulation of Quantum Mechanics, the one by Heisenberg, Born, Jordan and Dirac, rather that the later formulation by Schr\"odinger.   I believe that Schr\"odinger's contribution, effective in application, has been conceptually misleading.\footnote{\href{https://royalsocietypublishing.org/doi/10.1098/rsta.2017.0312}{CR, Space is blue and birds fly through it}, Phil. Trans. R. Soc. A (2018)37620170312.}   I have detailed this point from an historical perspective in a joint  paper with the historian of science John Heilbron.\footnote{ \href{https://doi.org/10.1140/epjh/s13129-023-00056-1}{J.L. Heilbron, CR, Matrix Mechanics Mis-Prized: Max Born's Belated Nobelization}, The European Physical Journal H, 48 (2023) 11.}

\item[Observers.] Importantly, no special role for ``observers" is assumed in RQM.  Variables take value with respect to any system.  What we commonly mean by ``observer" in our actual use of quantum theory \emph{does} assume special features of the system with respect to which variables take value, or, more accurately, of the setting in which this happens: these features are well described by the physical phenomenon of decoherence, and even more specifically by the phenomena investigated by quantum darwinism\footnote{Zurek, Wojciech Hubert (2003). Decoherence, einselection, and the quantum origins of the classical. Reviews of Modern Physics. 75 (3): 715–775.}.  These allow records to be formed and `stabilize' the values of certain variables.\footnote{Di Biagio, A., Rovelli, C. (2021). Stable Facts, Relative Facts. Foundations of Physics, 51, 30.}  These are the variables that we commonly call ``measured" in a ``quantum measurement".  The relational interpretation is meant to make sense of the theory also outside the special circumstances required by such complex ``measurements".  Physics happens also when there is no measuring apparatus and no records. 

\item[No elementarity and the Heisenberg cut.]  Contrary to what is carelessly assumed in several discussions, the systems that behave quantum mechanically are not just ``elementary" constituents of matter: any system does.   For instance, the angular momentum of a molecule is quantized: it only takes specific values, irrespectively of the fact that the molecule itself is a composite system formed by atoms, in turn formed by electrons and a nucleus, in turn formed by protons and neutrons, in turn composed by quarks, and so on.   All systems and all variables  display quantum properties: variables that involve many degrees of freedom are equally ``quantum" as the position of a single electron.  A ``system" is any part of Nature that admits a somewhat compact description and whose dynamics can be considered as isolated within some approximation.  Any such system is a quantum system.  The boundary of what we can call a quantum system is arbitrary. This is the freedom in choosing the Heisenberg cut: the boundary of the system described by quantum mechanics.  The quantum description of a system is relative to the choice of the Heisenberg cut.

\item[Discreteness.]  All quantum phenomena are characterized by the  constant $\hbar$, the quantum of action. $\hbar$ quantifies the discreteness, or ``granularity", of the interactions between the processes that make up reality: elementary interactions are discrete at the scale determined by $\hbar$. The amount of information that a system $B$ can acquire about a system $A$ by means of an elementary interaction between the two is limited by $\hbar$: in classical terms, the state of $A$ cannot be localized in its phase space in a region with phase-space volume smaller than $2\pi\hbar$ (per degree of freedom). This fact, captured by Heisenberg's uncertainty principle, expresses the elementary granularity of reality. The electromagnetic field, for instance, cannot act on a screen in a manner weaker than one discrete photon at a time. Classical continuity is traded for continuous probabilities of discrete $\hbar$-size quantum leaps.   Photons are not entities: they are the elementary discrete modes of interactions between the field and a screen.  The relational reading of quantum theory is the sole interpretation --as far as I can see-- that make sense of the fundamental granularity pervasive in all quantum phenomena. 

\item[Processes rather than things.] Like most of our knowledge of the world, the theory is not about entities: it is about processes. 

\item[Reality as mutual action.]  While classical mechanics describes reality as observed from an external point of view, QM describes it in terms of the interactions between one system and another. Stripped of anthropocentrism and instrumentalism, and hence the confusing notion of ``observer", this is a description of reality as interaction, described from within the network of relations itself.\footnote{Hans Halvorson traces this decisive shift to Kierkegaard's influence on Niels Bohr:  H Halvorson, The philosophy of science in Either-Or In: Kierkegaard's Either-Or, R Kemp, W Wietzke eds (Cambridge 2023) pp 153-170.} 

\end{description}

This relational understanding of quantum theory resolves traditional difficulties by adopting a perspectival view of reality.   Systems do not ``have" a state by themselves.  States are relative to something else.  Properties are relative to something else.  The notion of individual system, independently from anything else, is rather empty, since a system by itself has no (contingent) property. I shall address philosophical implications of this perspectivalism in the last Seminar. 

 \paragraph{Relational quantum mechanics and quantum gravity.} \label{q}

I can now get back to quantum gravity, and fill in some missing points, in light of the understanding of quantum mechanics that I just illustrated. The presentation of quantum gravity given in the first Seminar, may have left  a doubt. Why do we describe quantum spacetime by simply giving transition amplitudes between boundary surfaces?  What determines these surfaces? 

The answer to these questions is provided by the relational character of quantum mechanics emphasized above.  The boundary of the spacetime region described by a quantum gravity transition is a Heisenberg cut.   Quantum gravity is a description of processes.  These processes form spacetime.  They are relative to the spacetime region surrounding them.   The full structure is markedly perspectival and relational.  Quantum theory describes how systems manifest themselves to other systems.   In quantum gravity, systems are spacetime regions.    Quantum gravity describes how spacetime regions act on one another.    Spacetime regions are not qualitatively different from other quantum processes: they are just the peculiar form that gravitational field's processes take.

Hence, the  relationalism of quantum theory (systems are characterized by the way they affect one another), and the  relationalism of general relativity (spacetime is characterized by the way different spacetime regions connect to one another while localization is only relative to other dynamical entities,) are strictly interrelated.   In fact, they are ultimately the same, since locality implies that in order to affect one another two systems need to be adjacent and for two systems to be adjacent, only means that they can directly act on one another. 
 
Far from being ``incompatible", or ``profoundly different", as sometimes claimed, general relativity and quantum mechanics share a common relational core that makes them merge naturally.  They both are about the network of the interactions between processes.

\paragraph{Information.} The state $\psi$ of a system $A$ relative to another system $B$ codes the information that $B$ has acquired about $A$ in the course of previous interactions.  The precise meaning of `information' here is elementary. It is just physical correlation: the way a variables of $B$ can affect a third system predicts the way a variable of $A$ does. This is the basic definition of (mutual) information as  physical correlation that I will discuss later in the Seminar \ref{subject}. In the original 1996 paper on RQM, the hope was expressed that the formalism of QM could be derived from simple transparent physical postulates (in the way Special Relativity can), and two postulates were proposed: 
\begin{enumerate}[label=(\roman*)]
\item The maximal relevant information of a system is finite; and 
\item New relevant information can always be acquired.  
\end{enumerate}
Here `relevant' means non-redundant for prediction and the system is assumed to have a finite volume phase space.  The first postulate captures the discreteness quantified by $\hbar$. (The maximal info is determined by the phase space volume in $2\pi\hbar$ units.) While the second captures the openness implied by non-commutativity.  A rich literature of research programs aimed at deriving QM from postulates have since developed.  (Additional continuity requirements are needed to get the full quantum formalism from the two postulates above.)

  \paragraph{The classical world and the recovery of a shared reality.} \label{q}

Decoherence is the physical phenomenon for which when a system $S$ interacts with an `environment' $E$ and with a system $O$ that has no access to that environment, quantum interference effects can be strongly suppressed in the transition amplitudes of some variables of  $S$ relative to $O$.  Decoherence is well known for explaining why we generally do not see interference effects for macroscopic objects, and therefore we can drop the assumption of them being `in quantum superposition' (that is, display interference effects).  In RQM,  it allows us to see `stable' facts.\footnote{Di Biagio, CR: Stable Facts, Relative Facts,  Foundations of Physics, 51, 30 (2021), arXiv:2006.15543.} 
A group of physical systems $O_1... ,O_n$ sharing a common inaccessible environment determines a common decoherence domain. The classical world we live in, where we have stable facts that we can measure, register, remember, do science with, and construct historical accounts with, is such a decoherence domain.  Thanks to this, RQM  does not jeopardize our ability to do science and describe a stable world.  

\paragraph{Relativity iterates.} \label{q}

Still, if quantum theory is correct, then in principle the entire set of the observables that enjoy a common decoherence domain can be like Friend in Wigner's Friend scenario. That is, a ``super-observer" capable of  measuring these observables as well as the environment could still observe interference effects between the world they describe and some other configuration of the world.  In this (rather abstract) sense, the actual description of the world  that we give is always partial, relative to us (us as physical systems, not as thinking beings). Everything we see, could be a term in a quantum superposition, with respect to some ``super-observer".  This may be disconcerting: is there something about our contingent world that is not just relative? 

One way to try to tame the resulting lack of ground is to interpret relationality as relativism.   That is, to say that all that RQM requires is to recognize that the values of variables are relative.  What is absolute is then the fact that a variable has a value relative to a system. But this does not suffice, for the following subtle reason. To say that a variable of a system $S$ has a value relative to a system $O$ is to state something about the state of the coupled system formed by $S$ and $O$ together.   And this {\em in turn} is only true relatively to a system (possibly just $O$ itself), in quantum physics.    

This can be seen by considering the following example. Take the case that in the room where they are enclosed Friend interacts with $S$ and measures $S$ only if a quantum experiment gives him one outcome rather than another one.   In Wigner's account of the content of the room, there is a superposition between a state where $S$ has a value for Friend, and a state where it doesn't.  Hence the fact itself that $S$ has a value relative to Friend is only relative to something else: there is no non-relative fact of the matter about it.  

Now comes what I think is the most radical philosophical implication of all this.\footnote{See Timotheus Riedel: Relational Quantum Mechanics, quantum relativism, and the iteration of relativity, Studies in History and Philosophy of Science, 104 (2024) 109-118.}  Relativity iterates: that is, the above observation triggers an infinite regression that prevents us from making any meaningful absolute statement about what is the case.   If we pretend a proposition like {\em ``the fact X obtains"} to have an absolute meaning and not be relative to any physical system, RQM blocks this from being possible.  This is the radical perspectivism implied by RQM. 

It is not harmful in our practice of science, because what we describe is always the world relative to our shared decoherence domain in which we conduct our daily and scientific business.  But it is harmful for those looking for ultimate facts.   In the last Seminar I  come back to what seems to be general consequences of this observation. \\[1cm]

\section{Time}\label{t}

\epigraph {\em \hspace{6em}... The eternal current\\
sweeps all the ages with it through both kingdoms\\
forever and drowns their voices in both.}{Rilke, Duino Elegies, I}

%\epigraph {\em The time is out of joint.}{Hamlet, {V.}}

There are distinct lively debates around the notion of time.   It seems to me that they often mix up different questions.  Breaking the discussion into separate components is crucial because --as I will argue-- the notion of time is heavily layered: it is composed by a number of layers characterizing distinct aspects of temporality.   The layers are related, but distinct.   

A common mistake, I believe,  is to assume that there is a single entity, or structure, or notion, that captures everything we mean when we say ``time": the `essence' of Time.  There is nothing like that, and this is what generates the confusion.  To ask questions about ``the nature of time", without specifying which of these layers we are referring to, is meaningless.    It is talking about something that does not exists.

The word ``time", that is, is employed in a vast variety of different  ways, related to one another in different manners.\footnote{Cfr L Wittgenstein, Philosophical Investigations (from now on: `PI'), I.38, I.45 and elsewhere.} To make sense of it, we need to ``bring [it] back from [its] metaphysical use, to [its] daily use"\footnote{PI, I, 116.}.

\subsubsection{Layers}\label{tl}

In everyday life, we conceptualize reality as evolving in a unique global common time, where past events are fixed, present events are actual whether they are close or far from us,  the future is undetermined and can be affected by our choices.  This is a good way of conceptualizing time in our daily business, but it increasingly fails as we try to extend it to wider range of phenomena, where, instead, increasingly weaker notions of temporality are viable. Before addressing specific questions about time let me start by disentangling different layers. 
\begin{description}
\item[Experiential time:] This is our common sense of time: a notion strongly marked by the fact that we have short- and long-term memories, anticipations, and knowledge about other times, we deliberate, we have regrets, desires, narratives, fears, anxiety for the rapid flight of the minutes, the brevity of life.  The brain is (also) a mechanism for collecting memories of the past in order to use them continually to predict the future.  This happens across a wide spectrum of time scales, from fraction of seconds to centuries. The possibility of predicting something in the future obviously improves our chances of survival and, consequently, evolution has selected the neural structures that allow it.  We are the result of this selection. This being between past and future events is central to our mental structure.  This rich experience of time is grounded in the fact that the subject of experience (such as us) is itself a thermodynamic irreversible process, hence \emph{itself} marked by the entropic time orientation and the phenomena it gives rise to. I have discussed the  demarcation between the aspects of temporality pertinent to neuroscience and the aspects pertinent to physics in a joint paper with the neuroscientist Dean Buonomanno.\footnote{D.~Buonomano, CR. Bridging the neuroscience and physics of time, in {\em Time and Science, Volume 2: Life Sciences},  R. Lestienne and P. Harris eds., World Scientific 2023,  arXiv:2105.00540. In the paper, we are largely, but not totaly, in agreement.}
\item[Oriented time:] In strictly physical terms, this is the time described by the physics that includes the time orientation implicit in the Second Law of thermodynamics. (More precisely, the `minus first' law, that expresses the tendency of systems to equilibrate in one direction of time\footnote{Brown, H. R. and J. Uffink (2001). The origins of time-asymmetry in thermodynamics: The minus first law. Studies in History and Philosophy of Modern Physics 32, 525--538. WC Myrvold, The Science of $\Theta\Delta$, Foundations of Physics {50} (2020) 1219--1251.}). 
\item[Mechanical time:] When we take all the individual degrees of freedom of a sufficiently isolated system into account, and treat them on equal ground, we lose the \emph{orientation} of time: classical and quantum mechanics know no time orientation.   
\item[Special relativistic time:] When we consider phenomena where fast velocities are involved, we lose the \emph{uniqueness} of the time variable: different Lorentz times are on the same footing. 
\item[General relativistic (proper) time:] When strong gravitational field are involved, we lose the possibility of any physically meaningful \emph{global} time variable.\footnote{Approximate times are of course viable. An example is cosmological time, defined in the approximation where inhomogeneities are disregarded.} 
\item[Quantum gravity temporality:] When quantum gravitational phenomena cannot be neglected, we lose the local temporal structure provided by a Lorentzian geometry. No specific  variable plays any special ``temporal" role.  
\end{description}
The mathematics employed by different theories to describe temporality varies accordingly. Thermal, non-relativistic, relativistic and quantum gravitational theories employ different mathematical tools to treat temporality.   The relation between the different layers are generically well understood, in terms of approximations and idealizations. The challenge raised by each level to our common conceptualization of time are often under-appreciated.  Let me discuss them. 

\paragraph{Time in relativistic gravity.} 
There are two ways of formulating a mechanical theory.   The first does not generalize to relativistic gravity. The second does.\footnote{For an extended discussion, see CR, Quantum Gravity, op cit.}

The first is to interpret a mechanical theory as a theory for a set of variables $q_n=\{q_1,...,q_N\}$ evolving in time $t$.    The mathematics of the theory determines which functions $q_m(t)$ are physically realizable.  This knowledge allows us to infer the values of $q_m$ at some $t$'s if we have sufficient information about their values at some different $t$'s.

The second way to formulate \emph{the same} mechanics is to include the variable $t$ into the set of the other variables, and express the same information as above as a {\em relation} between the full set of resulting variables.  That is, call $x_a=\{x_o=t, x_n=q_n\}$ where $a=0,...,N$, and express the same information as above in the form $f\!(x_a)=0$.\footnote{To make this concrete, consider a pendulum.  Call $q$ the angle of the pendulum with the vertical and $t$ the reading of a clock.   The first way to express the dynamics is to give the generic dependency of $q$ on $t$. This is $q(t)=A\sin(t+\phi)$, for any two constants $A$ and $\phi$.  The second is to treat $q$ and $t$ on equal footing and express the dynamics by saying that the relation between the two must be of the form $f_{A,\phi}(t,q)=0$ where the possible functions $f_{A,\phi}(t,q)=q-A\sin(t+\phi)$, are labeled by any two variables $A$ and $\phi$.  In either case, $A$ and $\phi$ coordinatize the phase space.} 

The distinction between the two formulations of dynamics is subtle.   They are almost equivalent, but not completely.   The subtle difference is that the first singles out one variable as ``special": the $t$ variable. The second does not.   That is, the first includes a bit more information than the second: it specifies which of the variables is the time variable, by giving it a special role. 

In special relativistic dynamics, clocks move at different speed, depending on the Lorentz frame in which they are.  We have therefore a family of physically equivalent time variables, one per each possible Lorentz frame, related to one another by Lorentz transformations. It is then convenient to adopt the second formulation, in order not to privilege any specific Lorentz frame, and keep Lorentz covariance manifest. This is realized in all Lorentz covariant formulations of relativistic mechanics.\footnote{Where for instance a particle motion is described by the four functions $x^\mu(s)$ rather than the three functions $\vec x(t)$.} 

When moving to GR, adopting the second perspective is not an option: it is unavoidable.  GR is formulated as a theory that describes \emph{relative} evolution between variables, not how variables evolve with respect to a single privileged `temporal' variable.  

The physical reason of this necessity is simple. It can be appreciated as follows. Consider two equal clocks that have been synchronized.  Keep one on the table, throw the other upward, catch it when it falls back and put it back on the table near the other.   It is a fact of nature that the two clocks are then not synchronized any more. They display a time difference $\Delta t$ that depends on how high the clock has thrown.\footnote{Which one of the two clocks has measured the longer time? Careful: the answer is opposite in special and general relativity. Who is right?}  Questions: Which one of the two clocks measures  real time? Is the difference due to the fact that the first clock is evolving in the true time defined by the second, or viceversa?  

The right answer is that the questions are meaningless. There is no preferred time variable in GR and no preferred clock measuring it.   There are clocks measuring proper times along their own worldlines, and none of them, in general, defines a preferred time for the overall evolution. 

Our Galaxy and Andromeda are heading towards a collision: when they will meet, the times elapsed from the Big Bang will be different in the two galaxies.  None of the two will have any claim of being more of a ``true" time than the other.  This is a fact about nature.  It challenges our usual experience where all clocks define a common time, but our common experience is limited and cannot be trusted for informing us about universal structures of reality.  Our temporal intuition is blinded by the fact that in our lives relativistic time discrepancies are negligible. 

The way GR takes care of this natural fact is by not having a preferred time variable.   Dynamics is expressed as a relation between variables --as illustrated above--, and not  in terms of a preferred one.  That is, GR is formulated in terms of the second of the above formulations of dynamics.\footnote{Mathematically, this is implemented by general covariance.   To see how this works, let us get back to the pendulum and the clock, described by the variables $q$ and $t$, respectively.   A relation between them is given by a curve in the $(q,t)$ space. This curve can be described without choosing a preferred time variable by parametrizing it with an arbitrary continuous variable $\tau$ and giving the two functions $(q(\tau),t(\tau))$.   The mechanics of the pendulum can be formulated in this manner, treating reparametrizations of $\tau$ as a gauge.  The time coordinate of general relativity plays the same role as such a $\tau$ variable.}   Dynamics is a relation between variables.

\paragraph{Time in quantum gravity.}   \emph{Quantum} gravity is formulated like classical relativistic gravity.   In the example of the first Seminar, the dynamics is given as a \emph{relation} between  variables measured on initial and final boundaries.   If there are one or more `clocks' that display a value at the beginning and a value at the end of the process, any of these clocks can be equally taken as a ``time" variable.   The choice is not dictated by the theory, nor it has, in the content of the theory, any special physical significance or implication.  We can choose criteria under which we want to call `clock' a variable, but what we get is just what we put it: criteria for the use of a name.  In quantum gravity, as in classical GR, temporality is expressed by {\em relations} between the ensemble of the initial and final variables of a process.   In quantum gravity, as in classical GR, temporality is expressed by {\em relations} between the ensemble of the initial and final variables of a process.\footnote{Three comments may be useful to dispel confusion in the literature on time in quantum gravity:
\begin{itemize}
\item Confusion is  generated by the requirement of gauge invariance for the physical observables.  Observable quantities must be  invariant under gauges acting on both the system and the observer: they do not need to be  invariant under gauge transformations of the sole system. Quantities that can be measured in this sense are called ``partial observables" and studied in detail in \href{https://doi.org/10.1103/PhysRevD.65.124013}{CR, Partial observables}, PRD (2002) 65. There is nothing wrong in assigning a physical, operational meaning to these quantities.
\item
There is a vast literature on a supposed {\em disappearance of time} in quantum gravity, and on ways to `recover time'. All this is ill-conceived.  The disappearance of a preferred temporal variable is  there in \emph{classical} GR and has nothing to do with quantum phenomena.  Searching criteria for finding ``the right time variable" in a quantum formulation of gravity is nonsensical: there is no ``right time variable" in relativistic gravity.   Ideas about recovering time from \emph{quantum} interference or similar (as in the popular Page-Wootters formalism (DN Page, WK, Wootters, Evolution without evolution: Dynamics described by stationary observables. PRD27  (1983) 2885.) are seriously misleading: temporal relations show up in a quantum theory of gravity in the same manner as in the classical theory: as relations that the dynamics fix between observable quantities. 
\item There is widespread confusion regarding {\em unitarity}.  If there is a global time, namely a variable $t$ monotonic on all solutions of the equations of motion, such that the evolution in $t$ is a symmetry (if $t$-translation is a symmetry), then we expect this symmetry to be preserved in the quantum theory.  Then the state space carries a {\em unitary} representation of the time translations.  Much literature takes the existence of such unitary representation as a condition for probabilities to be well defined.   The mistake is to assume that the existence of a global time and symmetry under time translation are required for consistency.  The mistake, that is, is to promote our common intuition about how time to a consistency requirement about nature.  This is like promoting our intuition that the Earth is flat to consistency requirement for the theory.   Unitarity time evolution is not required by consistency.  
\end{itemize}
}

\paragraph{Neither Presentism nor Eternalism.} By ``present" we commonly mean the collection of the events where the time variable has the same value as here now.  In the light of classical GR, this notion ---a present arbitrarily extended in space, or the set of all the events of the universe that are ``happening now"--- is meaningless.  The same is true in quantum gravity.

That is, the question ``What is happening \emph{now} on a distant star?" is meaningless.   (The question ``What is happening \emph{here} in Beijing?" asked by somebody in Princeton, is equally meaningless.)    Along the history of a distant star there are events that are definitely in our past (we may have seen them) and events that are definitely in our future (somebody there has a chance to see what we are doing now).  In between, there is a long lapse of events that are neither past nor future to us now.  The duration of this lapse is the back-and-forth travel time of the fastest possible signal. All events along this time lapse are equivalent, as far as their temporal relation with the present here is concerned. 

The question ``What is happening now {\em in Paris}", on the contrary, \emph{is} meaningful for us in Princeton because in this case the time lapse is very short, given the short distance of Paris: the back and forth travel time of light from Paris to Princeton is of the order of milliseconds.   We do not resolve or care about milliseconds, and therefore we can safely say that any event in Paris happening during or along this millisecond time span is happening ``now".   For a distant star, the lapse is not millisecond: it can be centuries or billions of years.  The ``now" in this sense on that distant star can last for quite a long time. 

Because of these facts of nature, the notion of ``present state of the universe" is not appropriate to describes our universe.  The idea that this notion is needed to describe the world is called {\em Presentism}.   Presentism does not account for our real world.  Twentieth-century physics shows, in a way that seems unequivocal to me, that our world is not described well by Presentism: an objective universal present does not make sense, except in approximations.  

\pausa

From the failure of Presentism to describe our world, some philosophers (following an influential paper by Putnam\footnote{H.~Putnam, Time and Physical Geometry. Journal of Philosophy, 64  (1967) 240--247.}), and some physicists, have deduced that we should rather view reality as a four dimensional  {\em static} ``block" where nothing really happens and all events are equally existing now. Process, change, dynamics, evolution, temporality, are, according to this view, illusory, or in some other ways derivative.   This view is sometimes called {\em Eternalism}, or, better, `naive Eternalism', as   `Eternalism' is also employed by more articulated accounts of temporality.

I find this naive Block-Universe Eternalism unmotivated.  Physics is about processes, change, dynamics, evolution, temporality. Not about anything like a block, or about anything ``static".  Along each timelike worldine, temporality is defined with respect to the variable defined by any local clock tracking proper time. Locally, proper time behaves \emph{precisely} as the time of Newtonian mechanics. Therefore there is no ground for asserting that Newtonian physics describes evolution while general relativistic physics does not.   

The different individual times along different worldlines do not merge into a coherent global time, as they do in Newtonian physics.  Systems communicate but have no common timeline.  Relativistic temporality is incompatible with Presentism, but has no connotation of being ``static" or ``block", either. Nor it implies that change is illusory.  The models of GR do not describe extended four-dimensional objects which are not subject to change: they describe processes, or ways in which things change.\footnote{The point made in this paragraph is  developed in  \href{https://arxiv.org/abs/1910.02474}{CR, Neither Presentism nor Eternalism}, Foundations of Physics, 49(12), 1325-1335. arXiv:1910.02474.}  

\pausa

The reason some thinkers fall from (unworkable) Presentism into this (unmotivated) Block-Universe Eternalism is lack of imagination or capacity of conceptual flexibility.  Relativity does not deny temporality, processuality, change.   It requires us to \emph{adjust} these concepts to a deeper  understanding of the ways of nature. What Einstein  achieved is a revision in the way that we describe reality.  If we hang on to too simple-minded a notion of change, we get lost and fail to understand newly discovered physics. To hang onto a precise use of a concept is to close ones eyes to the possibility of a better understanding. 

A child grown in a family where males are aggressive and females  submissive get confused when going into the world and discovering that there are other ways. The child identifies maleness with aggressivity and deduces that non-aggressive men  are not  men.  (``That is not what \emph{I mean} by `man'!".)  Precisely in the same manner, a thinker used to identifying temporality with the uniqueness and globality of time (or some other contingent feature of our temporal experience)  goes into the world and seeing relativistic phenomena with no global temporality, deduces that this is not ``true" temporality.   (``That is not what \emph{I mean} by `time'!".) The intelligent child understands that there are different possible ways of being men. The intelligent thinker, understands that there are different possible ways of being dynamical and different ways in which temporality can be manifest.  Concept must adapt to knowledge, not vice versa. \\[.5cm]
 
\subsubsection{Flow}  \label{f}

%\epigraph {\em \hspace{6em}... The eternal current\\ sweeps all the ages with it through both kingdoms\\ forever and drowns their voices in both.}{Rilke, Duino Elegies, I}

Yet, our experience of temporality is far richer than just the perception of a relation between variables.  We perceive time as something that flows in a direction, it passes, it is Rilke's  {\em eternal current} that {\em draws all the ages along with it...}   in the epigraph of this Seminar. 

This ``flow"  may not appear easy to characterize (not by chance is a poet called to help here), but is vividly manifested by phenomena such as the fact that we remember the past\footnote{Augustine of Hippo, as far as we know, was the first to draw attention to the fact that we have experience of the flow of time only because we have memory.  His page in the Confessions on how we perceive a musical  theme only because we have memory is memorable.} but not  the future; we can decide the future but not  the past; effects (a ring of expanding ripples) follow, never precede, their cause (a stone falling into the pond); a measuring instrument records past measurements, not future ones, and our very thinking \emph{unfolds} in time, from past to future.  

These and similar phenomena --I would like to claim-- are not {\em manifestations} of the flow of time: they {\em are} what we mean by the flow of time.  To understand why time flows we must understand these phenomena. 

These \emph{directed} aspects of temporality appears to be absent in the equations of GR, which are invariant under time reversal.  But they appear to be equally absent in the equations of Newtonian classical mechanics.  The fact that Newton gives time a special name (``$t$") to a variable and treats it as the independent variable, is not an explanation or an account of the directed flow. Naming a phenomenon does not account for it.  

So, what is this directed ``flow"? What sources the vivid oriented phenomena characterizing it?

\pausa

The answer to this question is that accounting for the oriented aspects of temporality is not to be found in the fundamental physical laws alone---be these mechanics, electromagnetism, quantum theory, GR, or quantum gravity.  The answer has to be found in the phenomena that we call thermodynamical. 

This can be seen by simply asking the question whether there is any instrument that measures the \emph{direction} of time flow.   The answer is of course yes: there are plenty of instruments that measure the arrow of time.   A candle does.  Any irreversible phenomenon does, and irreversible phenomena are ubiquitous.  In fact, even all our clocks do, because the time they display increases monotonically, and this is because any clock includes some amount of irreversibility.    All  systems that detect the arrow of time do so by virtue of some irreversible phenomenon, namely by tracking some form of \emph{dissipation}, for example of work into heat. In turn, such dissipation is measured by what we call an entropy production.   

The functioning of our brain is highly dissipative, hence there is no surprise that our thinking, which is the way we call the activity of our physical brain, is also markedly time oriented, and easily detects a direction of time. 

It is sometimes claimed that the flow of time our conscious experiences is not captured by physics and cannot be measured by instruments. This is nonsense. Our brain detects the local entropy production precisely like all the other instruments mentioned above do.  I find the idea that we could detect something that cannot be detected by a physical devise to be funny and silly.

The question here is not a hypothetical discrepancy between the flow of time perceived by our consciousness and the world described by physics; it is rather the puzzling discrepancy between the ubiquitous time-oriented irreversibility of the phenomena that we see and the time reversal invariance of fundamental physics. 

Irreversible phenomena show up in the statistical treatment of systems with a large number of  (``microscopic") variables, when we focus on a small (``macroscopic") subset of these.  This is the domain of statistical mechanics.  Dissipation happens when energy flows from the second to the first.  Oriented temporality is a property of the macro-variables, hence a feature of an approximate (``blurred") macroscopic account.

All phenomena related to the directional flow of time (memory, traces, agency, choices, time oriented causation...) regard \emph{macroscopic} variables and are characterized by the existence of dissipation: energy flows from macroscopic variables to microscopic ones.\footnote{See D Albert, Time and Chance (Harvard University Press, Cambridge 2000). WC Myrvold, The Science of $\Theta\Delta$, Foundations of Physics {50} (2020) 1219--1251. \href{http://philsci-archive.pitt.edu/17328/.}{B Loewer, The Consequence Argument Meets the Mentaculus} (2020).} 

This is counterintuitive. It is perhaps the most astonishing discovery of modern physics.  There are some thinkers, including philosophers, that do not accept this discovery. They find this idea inconceivable.  They hang on to the idea of a \emph{fundamental} orientation of time. 

I believe that these thinkers are like those who hang on the idea that it is impossible to conceive that the Earth is actually rotating.  The second hang on to the idea that the ground does not move by the very nature of motion, and reality would be unconceivable otherwise.  The first hang on to the idea that our naive intuition about time captures something necessary, without wich reality would be inconceivable.   The error is to mistake a (vivid) feature of the manifest image for something universal.  Nature is wider and smarter than our prejudices.  Once again, the mistake is to put concepts before knowledge.  What is not ``conceivable" for some, at some historical junction, becomes conceivable for others, or at another historical junction, because concepts evolve.  To be ``conceivable" means to be graspable by means of \emph{available} concepts. 

In the next paragraphs I illustrate \emph{how} a (contingent) gradient of entropy gives rise to the vivid sense of the flow of time.   The way this happens is far from obvious; it is especially subtle because it has two sides, one regarding observed phenomena, the other regarding the subject of knowledge itself. 

\paragraph{The phenomena that constitute the flow of time.}

Dissipation is the source of the time-orientation of all the ubiquitous phenomena in our universe that are time oriented.   The quantity that traces dissipation, the flow of energy between macro and micro variables is entropy.\footnote{When temperature is defined, entropy can be thought as the difference between the total energy of a system and the energy accessible with the available macro-variables (the ``free energy"), divided by the temperature.}  Roughly, we call \emph{work} the energy exchanged via macro-variables and  \emph{heat} the energy stored into micro-variables and irretrievable by only acting on macro-variables.  The distinction between heat and work is therefore dependent on the distinction between macro-variables and micro-variables.  More precisely, since the micro-variables are independent of  this distinction, it depends on what we count as macro-variables.   I discuss what is precisely meant by macro-variables later on. For the moment, let us just take this distinction for given.  The most common form of \emph{dissipation} is the transfer (the ``loss") of energy from work into heat.  All time oriented phenomena are dissipative.   Here is a discussion of particularly relevant cases, on which I have worked in recent years. 
\begin{description}

\item[Traces.] It is a fact that the present is full of features that we can interpret as traces of the past, and there are no analogous traces of the future.  A crater on the moon is the trace of the impact of a meteorite, a step in the sand is the trace of a step of a man, a photo is the trace of the appearance of somebody younger, outgoing concentric waves on a pond are  traces of the falling of a stone into the pond.  In  all these cases the time orientation is due to dissipation: the meteorite, the step, the light impacting the film of the photo, the stone falling in the pond, carry an energy higher than the thermal energy of the target, which then gets slowly (because of the long thermalization times) dissipated into the target.  I have given a detailed analysis of the thermodynamics of this mechanism in the paper {\em Memory and Entropy}\footnote{\href{https://www.mdpi.com/1099-4300/24/8/1022}{CR, Memory and Entropy}, Entropy 2022, 24(8), 1022, arXiv:2003.06687.}.  The simplest scheme for understanding traces is the following. Consider a system $H$ with high temperature $T_H$, interacting occasionally with a collection of systems $L_1, ... L_N$ with lower temperature $T_L<T_H$.  Let $\tau_L$ be the time it takes the low temperature systems to equilibrate among themselves and $\tau_H\gg \tau_L$ the time it takes the full set of systems to equilibrate. Consider an occasional event $\cal E$: an interaction between $H$ and one of the $L_i$'s, say $L_1\equiv {\cal T}$ (for `Trace'). By the temperature difference, an amount of energy $Q$ is transmitted from $H$ to ${\cal T}$, which then remains at a higher temperature than the rest of the $L_i$'s, during a time lapse $t<\tau_L$. The system ${\cal T}$, being at higher temperature than the others $L_i$'s is a \emph{trace} of the past event $\cal E$. The formation of the trace requires an entropy increase $\Delta S=Q/T_L-Q/T_H$: the information in the trace is paid for by this increase of entropy.    This is general thermodynamic structure of trace formation. It requires dissipation. 

The functioning of our brain, in particular, relies on storing traces of the past, in the form of memories and acquired skills.  Memory is a particular case of trace.   A remarkable aspect of the mechanism that creates traces, described above, is that it can be seen as a way to promote low entropy into macroscopic information, in particular memory.  It can be shown that in order for a trace to fix an amount of information $I$ about the past, a corresponding amount of entropy $S\sim k I$ must be raised.  Accordingly, the amount of information $I$ fixed in a trace can be bounded by thermodynamical relations. For instance, if a trace is fixed by moving a small amount of energy from a more energetic environment at temperature $T_e$ to a memory system at temperature $T_m$, the amount of information captured in the trace satisfies 
\be
I < 
\Delta S/k = \frac{C(T_e-T_m)^2}{kT_eT_m} (1 - e^{-\tau_m /\tau_{e}}). 
\ee
where $C$ is the heat capacity of the memory and $\tau_m$ and $\tau_{e}$ are the thermalization times for the memory to dissipate and the memory-environment to thermalize.  The interest of relations like this is that they show that the entire information contained in our memories, books, our entire civilization, in the biosphere, can be seen as sourced in the initial low entropy of the universe. 
 
\item[Measurements.] Nothing can be measured without dissipation.  This is discussed already in the 1950s by Reichenbach, in ``The Direction of Time"\footnote{H. Reichenbach, The Direction of Time, Dover 1956.}.  The proof is  simple. After the measurement, we expect the record to be appropriately correlated with the measured system; in a purely mechanical context without dissipation, this expectation could not be justified, because any joint configuration is equally likely.  It is the low entropy of the past that allows measurement and recording of measured quantities.  The way this works is as a special case of the general case of traces: a record is a trace of an interaction between a system and an apparatus, because false records are probabilistically disfavoured. Measurements require dissipation and can only happen in an entropy gradient. 

An important consequence: any knowledge of the world requires measuring it, measuring it requires dissipation, dissipation is losing information, in the sense that the memory is stored in macroscopic variables only.   Hence knowledge is \emph{necessarily} incomplete, on physical grounds, by its very nature. 

\item[Causation.]  ``Cause" is an expression with a variety of meanings.  (Aristotle famously lists four notions of cause\footnote{A. Falcon, Aristotle on Causality, {\em Stanford Encyclopedia of Philosophy 2-11}.} and Buddhist philosophy six\footnote{B. Dhammajoti, Sarv\=astiv\=ada Abhidharma, Centre of Buddhist Studies, The University of Hong Kong 2009.}.) The notion of cause I consider here is the time-oriented one where an event causes an effect that happens later in time.\footnote{Notice the proximity between this notion of causation and the notion of trace discussed above:  a crater on the moon is caused by the impact of a meteorite, a step in the sand is caused by a step of a man, and so on. The time orientation of all these processes is always due to dissipation: the meteorite, the step, the light impacting the film of the photo, the stone falling in the pond, all carry energy higher than their target and this energy gets partially (but only partially due to long thermalization times) dissipated into the target.}

Consider a process $\cal E$ where a system $A$ interacts at time $t=0$ with a system $S$ and this `causes a modification $B$ in $S$'.  What does this mean?

The best understanding of causation that we have, capable of distinguishing causation from simple correlation, is in interventionist terms.\footnote{\href{https://plato.stanford.edu/entries/causal-models/}{C Hitchcock, Causal Models}, in The Stanford Encyclopedia of Philosophy, EN. Zalta and U Nodelman eds.}  $A$ causes $B$ means that if the interaction does not happen, and {\em in the past} the system $S$ is in the same state as in $\cal E$, then the dynamical laws governing the behavior of $S$ alone give a different process, say ${\cal E}_f$, with a different \emph{future} for $S$ (if the stone had not fallen, the pond would not have ripples).   

This interventionist understanding of causation is clear, but what gives it its time orientation? Why, that is, do we consider the alternative story ${\cal E}_f$, and not the equally possible, time reversed, alternative? Namely: if the interaction does not happen, and {\em in the future} the system $S$ is in the same state as in $\cal E$, then the dynamical laws governing the behaviour of $S$ alone imply a different process ${\cal E}_p$ with a different  \emph{past} for $S$ (if the stone had not fallen, the ripples had to be already there in the past.) Why, that is, do we {\em assume by fiat} that interventions modify the future and not the past?  After all, it is perfectly possible, according to \emph{mechanical} laws, for outgoing concentric ripples to be produced by previous concentric ripples of the water, without any stone falling at $t=0$.  It suffices to time reverse the dynamics of the (dissipating!) ripples and we have a possible \emph{mechanical} evolution giving rises to similar ripples: same future and a different past.  Why then do we say that the falling stone changes the future, and we do not say that it changes the past?

The catch, and therefore the answer, is that  ${\cal E}_p$ implies a time inversion of dissipation. That is, ${\cal E}_f$ and  ${\cal E}_p$ are equally viable mechanically, but not equally viable thermodynamically:  the second contains time reversed dissipation, which we never observe in nature.  The first is `closer to actuality' as David Lewis puts it.\footnote{\href{https://joelvelasco.net/teaching/5330(fall2013)/lewis79-CFTimesArrow.pdf}{D Lewis Counterfactual Dependence and Time's Arrow}. Nous, 13 (1979). 455--476.}  I find Lewis' analysis of the source of the time arrow in causation exactly right.  But Lewis concludes: ``I regret that I do not know how to connect the several asymmetries I have discussed and the famous asymmetry of entropy.'' This is precisely what is done in the paragraph above. That is, it is dissipation that makes the difference: the alternative world with the same past is consistent with dissipation, while the alternative world with the same future is not.  It is dissipation that underpins the time orientation of causal talking.  This is why causation is absent in fundamental mechanics\footnote{B Russell, On the Notion of Cause, Proceedings of the Aristotelian Society 13 (1913) 1--26.}. More precisely, it is dissipation that determines the correlation patterns in the world that we read as causation, by rendering one of the two alternative stories `closer to actuality'. I have discussed this point in detail in \emph{How Oriented Causation Is Rooted into Thermodynamics}.\footnote{CR, How Oriented Causation Is Rooted into Thermodynamics. Philosophy of Physics 1(1) (2023) 11, 1--14.  arXiv:2211.00888. See also CR, Agency in Physics, in {\em Experience, abstraction and the scientific image of the world. Festschrift for Vincenzo Fano}, Franco Angeli editore, Milano 2021, arXiv:2007.05300 and  \href{http://philsci-archive.pitt.edu/20148/}{CR, Back to Reichenbach}. To appear in Journal for General Philosophy of Science. See also \href{http://arxiv.org/abs/2208.02721}{E Adlam, Is There Causation in Fundamental Physics? New Insights from Process Matrices and Quantum Causal Modelling},  arXiv:2208.0272.} (I will come back to causation, with one further observation.)

\item[Thermal history of the universe.] %In the gas case, the pressure difference between the two chambers can produce work. After thermalization, the pressure reaches the equilibrium between the two chambers and this potential source of work, namely this ``free energy", is ``dissipated" into work.   All these notions are dependent on the macro-micro distinction. 

In mechanical terms, the history of our universe is generic: one out of the very many allowed by the equations of motion.  The same history running backward in time would be an equally generic solution.  But if a distinction between macro and micro is established, a time orientation running through the entire history of the universe shows up: the same history running backward in time violates the Second Law, more or less everywhere.  

A key and often neglected fact to consider in making sense of temporal phenomena is that thermalization times are commonly very long.  The Sun has been burning for a few billions years, the Earth's core is cooling since its birth, and yet these bodies are still far from equilibrium.   The universe we see is something like 14 billions years old and far from having reached any form of equilibrium.  We find systems in equilibrium only when they happen to remain particularly well isolated, and even in this case, almost always the state of the system is metastable at best, and not stable: the system is not at maximal entropy.  For instance, any object, unless it is a block of iron, could raise its entropy by nuclear fusion or fission. 

Let's take a rapid look at the thermal history of the universe around us. Cosmology provides a credible account of the large scale happenings of the last 13 or 14 billions years in the region of reality our telescopes have access to. Full GR is not needed to make sense of this account, at least in its main outline. Because of the approximate large-scale homogeneity and isotropy, and the slow relative velocity of the galaxies, a simplified version of the theory suffices, where the complexity of gravity is reduced to just  two phenomena: Newtonian attraction, and the existence of a degree of freedom, called the scale factor, which gives the size of the physical space  available to a (roughly) fixed amount of matter.  The scale factor can be thought of as the varying volume of the universe, as if the universe was contained into a (3-)sphere with changing volume.  Strictly speaking, there is no common time in the universe ---every trajectory has its own temporality---, but the universe as a whole is homogeneous enough to admit an approximate description in terms of a common time, the cosmic time. 

Roughly 14 billions years ago in this time, the universe was in a macroscopic state that can be (approximately) described as matter in thermal equilibrium highly compressed in a very small volume.   Matter was in thermal equilibrium, but not the whole system: the volume degree of freedom was strongly out of equilibrium with respect to matter.  Hence, the past low entropy is nearly \emph{entirely} due to the smallness of the scale factor.\footnote{Homogeneity of the other gravitational degrees of freedom --emphasized by Roger Penrose-- is there is principle as well, but in practice its role has been minor in the real universe, in comparison.}  The early rapid expansion threw matter out of equilibrium.  In particular, at nucleosynthesis protons and neutrons got trapped into Hydrogen and Helium in a ratio which is not the equilibrium one, because the Hydrogen/Helium thermalization time became rapidly too long, with the lowering of the temperature of the universe.  These events generated an enormous reservoir of free energy (namely disequilibrium) in the form of atomic or molecular Hydrogen present everywhere in the Universe.  This free energy fuels most observed phenomena.   Gravitational instability triggered star formation, namely channels for the entropy to grow rapidly by fusing Hydrogen into Helium.  Vast amounts of free energy so released in our Sun flood the Earth and provide the free energy that fuels the biosphere.   

\item[Biology.]     The processes forming the Earth's biosphere are markedly time oriented. In his (otherwise) spectacularly insightful and prescient text {\em What is Life?}, Erwin Schr\"odinger  ---however---  disseminated a misleading idea: the idea that something very peculiar has to happen with thermodynamics for life to exist.  Organisms, according to this idea, need somehow to find ways to lower their internal entropy in order to have and maintain structure. I think that this idea is wrong and misleading. Life is not an anti-entropic phenomenon in any (even weak) sense, as often wrongly assumed: on the contrary, it is a process for entropy to grow faster and free energy to be degraded  into heat faster, like in a burning candle.\footnote{More details on the thermal history if the universe can be found in CR, Where was past low-entropy?,   Entropy 21 (2019) 466.  arXiv:1812.03578.}

The intuition behind the wrong idea is that `structure is order 'and `dissipation destroys order': hence dissipation cannot generate order.  But this intuition is incorrect. Left alone, mixed water and oil \emph{separate}, creating order at the scale of the glass.  The beautiful complex structures of snow crystals form {\em spontaneously} in an isolated box.  Disorderly bouncing balls in a isolated room all end up orderly on the floor, thanks to dissipation. And so on. All these phenomena happen by increasing entropy locally, not by decreasing it.  The idea that any formation of order is  disfavoured by entropy increase is flawed and is still misleading even many acute thinkers.   

What  goes on in all the cases above is that order at some scale (the glass, the snow flake, the position of the balls...) increases the phase space volume available to the system.  For instance, the bouncing balls dissipate their kinetic and potential energy into thermal energy, and this enlarges the  number of accessible micro-states for the system. Microscopic disorder largely pays the bill for novel macroscopic or mesoscopic order.  Structure formation is commonly \emph{favored}, not hindered by the drive towards dissipation.  This is why we see structure everywhere in nature, and not only in biology: from beaches with  stones all of the same dimension, to stellar systems dotting the galaxies.  

Earth is next to a fiercely burning object, a star.  A star is a channel discharging the vast thermodynamic imbalance frozen in the Hydrogen/Helium imbalance at nucleosynthesis.  It is therefore a fierce generator of free energy, which invests  Earth in the form of high energy photons.  Earth re-emits this energy towards the dark sky in the form twenty times as many lower energy photons, thus it finds itself with an overflow of free energy.    Thermodynamically, the biosphere is a byproduct  of this overflow.   It is a way (among many possible) for dissipation to happen.   In fact, what living organisms mostly do is to dissipate energy, facilitating the growth of entropy. They increase entropy growth, far from somehow resisting it.    If we kill one of two living creatures in two equal isolated rooms, the entropy  grows \emph{more} in the room with the living creature, for the same reason for which the entropy of an isolated room grows more if there is a burning candle than if there is a candle not burning.\footnote{For more details on all this, see K Jeffery, R Pollack, CR , On the statistical mechanics of life: Schr\"odinger revisited,  Entropy 21 (2019) 1211, arXiv:1908.08374.} The biosphere is a form of rapid dissipation, not a form of resistance to dissipation. 

The individual biochemical reactions that constitute life happen because they raise entropy (otherwise they would not happen).  The organisms acquire free energy via nutrition and dissipate it in metabolism.  At a larger scale, evolution itself is time oriented. It can be seen as a secular exploration of increasingly large phase space regions corresponding to  new structures and processes (phase space can be seen as a list of possible states but also as a list of possible solutions of the dynamical equation, namely different processes) searching for ones that better dissipate energy.   The time orientation of biological phenomena is rooted in the thermodynamic gradient, like all other oriented processes.  

\item[Clocks.] I close this list with one further simple observation, perhaps surprising: dissipation is needed even for clocks. No clock could work without dissipation.  Dissipation is obvious in a sandglass, that degrades potential energy into heat, and similar devices. It is less obvious is clocks based on oscillations, like a pendulum clock, where dissipation seems to be just an hindrance; but it is actually equally essential. A pendulum clock needs an  escapement mechanism to convert an (ideally frictionless) periodic motion, which knows no time direction, into a directional counting device.  The escapement needs dissipation to work: at thermal equilibrium, the hands of a mechanical clock would oscillate back and forth, rather than move steadily.  A beautiful lecture by Richard Feynman\footnote{\href{https://liberalarts.org.uk/richard-feynman-entertaining-lecture-on-time/}{R Feynman, The Messenger Lectures: The Distinction of Past and Future}, given at Cornell in 1964.} illustrates the detailed mechanics of this dissipation needed for a clock to work. (That lecture is also a feast of enlightening observations relevant for the philosophy of science.)

\end{description}

All the cases above converge to the same conclusion: the phenomena that nurture our vivid sense of the orientation of time stem from the thermodynamic orientation of dissipation, which is to say, by the entropy gradient. The flowing of time is the name we give to our relation to this gradient.

\subsubsection{Openness}  \label{ts}

There is one last vivid phenomenon that I have not yet discussed. It plays a particular role, and it opens a central topic for what follows.   This phenomenon is the openness of the future. 

The fact that the past is fixed, determined, while the future is open, undetermined, is perhaps for us \emph{the} main difference between the two directions of time.  In spite of its evidence, it is not very simple to say what we exactly mean when we say that the past if fixed and the future is open.  To some extent, we refer to the fact that we know much more about the past than about the future, and different macroscopic futures can evolve from the same macroscopic past\footnote{See \href{http://philsci-archive.pitt.edu/17328/}{B Loewer, The Consequence Argument Meets the Mentaculus}.}; but this is not what we actually mean by the openness of the future, because the future could be fixed and just not known.

What we mean about the openness of the future is that we can change it, at our will.   More precisely that what is going to happen in the future can depend on us.  This is the direct source of our evidence for the openness of the future. But what, in turns, does the fact that the future depends on us actually mean?  After all, we are ourselves parts of nature and we follow the same (possibly very complex) patterns as the rest of nature, whether these are probabilistic or deterministic. 

\paragraph{Playing chess.} To begin approaching the answer to this question, consider the following concrete physical case.  Consider a laptop computer that runs a program that plays chess.   Assume that the battery is well charged, the laptop functions well and does not break during this run.   Then we can assume that its behavior is completely deterministic: it is completely predictable, on the basis of the initial state of the chess pieces and the moves of the chess opponent. What happens in a game is that the program inputs the positions of the pieces and moves then accordingly, following moves of the opponent.   The computer does not have enough memory for having stored all possible games, and therefore computes its move more or less like a human chess player does: analyzing various possibilities and considering the different evolutions of the game that would follow if it choses one move or another.   On the basis of the results of this computation, and a set of weights it has learned, and stored, the computer choses a move.  There is nothing mysterious or controversial in all this.  Chess-playing apps are in many computers. 

Now consider the following question.   Is the future open, before the laptop makes a move?  Or is rather the future move of the laptop `already determined'?   

A moment of refection shows that the question is ill posed.  Under our assumptions, the entire complicated calculation performed by the computer is deterministic, hence there is a unique possible outcome it can have.  In this sense, the move is `already determined'.   But there is another sense in which the move is not determined until the laptop computes it, since the process of computation is precisely what  leads to the determination of which move to choose.  Nobody can deny that there is a also sense in which it is the calculation that determines the move: without that calculation, that is, the move would not happen. 

Say that we choose to say that the outcome of the move is `already determined'.  Then the logic of the computation becomes incomprehensible: if the outcome is already determined, what is the point of computing many possible future alternatives, evaluating the outcomes, and choosing among these?  If the future is already determined, there is no point in analyzing alternatives.  But if the laptop does not analyze alternative futures, it does not make a move and the determined outcome does not obtain, so then, why did we say it was determined? 

The subtlety is that here the dynamical path through which the future comes about involves a complex process that can be interpreted as an analysis of possible alternative futures.  Within this logic, a branching tree of future possibilities plays an essential role. 

\paragraph{Deliberation.} Much of the same goes on when a human being makes a choice.  Not because a human being is a deterministic system (most likely it is not, at the very least because of quantum physics), but because any  deliberation process can be interpreted as an analysis of possible alternative futures.  Human choices are determined by a highly variable combination of rational analysis, instinctual spontaneity, brain's arbitrage of different drives. It involves electric, chemical, thermodynamic, statistical and possible quantum phenomena, in ways we are far from clearly understanding.  But this is irrelevant for our sense of the openness of the future.   What is relevant is that deliberation itself involves contemplating different futures, and this is seeing future as open.  There is no contradiction whatsoever between this openness and the cogency of (deterministic or probabilistic) physical laws, as there is no contradiction between saying that the laptop behavior is deterministic and yet the laptop choses between  alternatives. 

The subtle point here is perspective. We \emph{are} deliberating agents.  Hence from our perspective, the future is open.  It is really so, and the future does depend on what we decide: it depends on the deliberating process that we are.  It is F.P.~Ramsey who pointed out the crucial fact\footnote{F.P. Ramsey,  General Propositions and Causality. In {\em Foundations: Essays in Philosophy, Logic, Mathematics and Economics} D. H. Mellor editor, pp. 133‚Äì151 (Routledge and Kegan 1978).} that for us the ultimate contingency is \emph{our own} agency. 

In an elementary physical sense, an ``agent" is any system that we regard as affecting another affected system.  When an agent (in this sense) is a system involved in a complex deliberation process, what happens in the future depends on the deliberating process, which takes possible alternative outcomes into account.  {\em From the perspective of such agent}, the future is open in a very precise sense: the calculation is formulated and conducted as an analysis of distinct {\em possible} alternative futures.    

This is not in conflict with determinism, as the example of the deterministic computer program playing chess shows.  The behaviour of the computer program is here fully deterministic, and yet the program functions on the premise that different possible futures can be analyzed and one of them is chosen as the suitable one by the (deterministic) program itself.  There is no contradiction, because the future is  determined by the past, but it is so {\em through} the mediation of the process going on in the computer.  This process is properly understood as analyzing possible alternative futures. It is correct to say that the future is determined by the past as well as saying that it is determined by what happens in the agent. Dependence is transitive, $C$ can depend on $A$, without prejudging the fact that $C$ depends on $B$ that depends on $A$.  

The apparent contradiction between our sense of agency and the possibility of an underlying determinism is a consequence of a conceptual mistake: imagining that the mental process of the decision is outside ---distinct from--- the normal course of nature.   It is not.   The knowledge of the subject and its active role in bringing about the future is part of the natural events, not external to these.\footnote{There is nothing really new in these ideas. They are perfectly clear already in \href{http://en.wikisource.org/wiki/Ethics_(Spinoza)/Part_3}{B Spinoza, Ethics, III}, 2 and Scolio.}

\paragraph{Ismael's interference.} The way our agency determines our sense of openness of the future has been acutely investigated by Jenann Ismael\footnote{Jenann Ismael, Jenann, The Open Universe: Totality, Self-reference and Time, {\em Australasian Philosophical Review} to appear.}, who has pointed out a phenomenon which she calls ``interference", at the root of this issue.  Deliberation requires knowledge and depends on knowledge.  Knowledge therefore affects the future.  It can even do so undermining a prediction, for instance an agent can choose to act in the opposite way as what it has been predicted to do.   

What eliminates the apparent contradiction is realizing that knowledge is not in a realm outside physics. Any form of knowledge held by a subject must be implemented as a physical configuration of the subject itself. Knowledge can affect the future because different knowledge corresponds to different physical states.  Since the possibility of undermining a true prediction contradicts logic, hence is impossible, the only possibility is that freedom to undermine a prediction necessarily implies that the prediction was unreliable and therefore in a deterministic context knowledge must necessarily be incomplete.  The agent deliberates, and chooses the future accordingly, therefore the future is necessarily unpredictable, hence open {on the basis of the knowledge available to the deliberating agent}.  The future is genuinely open to us precisely because we can choose it. And we can because the way that our future comes about depends on what we do in the phase of deliberation.\footnote{See also CR, Knowledge is embodied. Comment on Jenann Ismael's The Open Universe: Totality, Self-reference and Time, to appear.}

\paragraph{Causation, II.} These considerations help us understand why causation is so prominent in our way of interpreting natural processes.  They show {\em in which sense} it is the future and not the past to be determined by the action of an agent that considers alterative futures.    As Huw Price puts it, the \emph{interesting} question isn't what causation ``is", but rather (from a pragmatist perspective) how come \emph{we} think and talk in causal terms.\footnote{\href{https://philpapers.org/rec/PRICP}{Huw Price, Causal perspectivalism}, in {\em Causation, Physics, and the Constitution of Reality: Russell's Republic Revisited} 2007, pp.~250--292.  Time for Pragmatism,  in {\em Neo-pragmatism}, J.~Gert, ed. Oxford University Press 2003.}   We do so, because this is the logic in which we function: analyzing possible futures and acting accordingly.  The future is open for us precisely in the sense that we examine options and act to choose among them (and this, as the example of the computer playing chess shows, is compatible with determinism.)

Notice that here is the process that defines {\em the agent} to be markedly time oriented.  And this is of course not surprising, because  an agent (like us) that collects information, thinks, analyses and deliberates, is \emph{itself} a time-oriented macroscopic phenomenon, obviously.  As mentioned, the biosphere is a time oriented macroscopic phenomenon.  Thinking is a process that happens in our brain, part of the biosphere, and which is very dissipative and time-irreversible.  

Notice that this concerns the time orientation of the {\em subject} of knowledge, rather than the observed phenomenon.  \emph{This time orientation of the process of knowing is the most direct source of the vivid sense or reality's time orientation that we experience: we ourselves are time-oriented phenomena}. 

It is not just the phenomena that surround us that are time oriented: more importantly, we are ourselves time-oriented phenomena. But it is not just ourself that are time-oriented either: macroscopic phenomena \emph{are} time oriented ---entropy grows, irrespective of us---. 

What has made the biosphere and us, what generates the patterns providing the opportunities that evolution has exploited to determine the way we think, is the existence of an entropy gradient in our environment. But we fail to fully understand the relevance of causation, the openness of the future, and the profoundly intuitive aspect of the orientation of time, unless we trace it back to the fact that our very thinking ---in fact, our very living--- is a time oriented process. 

To put it vividly: at thermal equilibrium there is no distinction between past and future, but not so much because there are no irreversible phenomena: the full reason is that at thermal equilibrium our own functioning, our acquiring of knowledge, would not work.   We can imagine a universe at thermal equilibrium and a \emph{separated} observer seeing it fluctuating in a time that is still flowing ahead; but this imagination represents an universe at equilibrium and a separated observer that still dissipates into a gradient of entropy. Hence the time orientation that this agent would experience would still be determined by dissipation: its own, not the one of the observed thermal bath. 

Therefore, causation is rooted in the thermodynamic gradient through two distinct ways: because of the time orientation of the dissipation around us, causation is something that can be read {\em in the world}. But it is because of our own time orientation and because {\em we} are deliberating agents, that causation is something that {\em we} read on the world.\footnote{See also \href{https://framephys.org/causal-perspectivalism-conference-june-2022/}{J.Ismael, It's not what you look at, it's what you see}, in Causal Perspectivalism Conference‚ June 13-14 2022.}

\pausa

Before closing this Seminar devoted to the notion of time, I add a few comments on the connection of its second half, where I have shown that the orientation of time is due to the gradient of the entropy in a given time variable, with the first, where I have shown that at the most general level we have currently access, which includes relativistic gravity, there is no preferred time variable.  

To clarify this connection, we would need a well developed thermodynamics and statistical mechanics of the gravitational field. But we do not have such a theory, not even at the classical level\footnote{We do have a well developed and credible theory of thermodynamic and statistical mechanics of matter on a curved geometry. But this does include the effects of the thermalization of the gravitational degrees of freedom.  Even at this simpler level, surprising and intriguing things happen, for instance in a static equilibrium situation the temperature measured by a thermometer is not constant! See the discussion in H.M. Haggard, CR, Death and resurrection of the zeroth principle of thermodynamics, IJMP D  22 (2013) 1342007, arXiv:1302.0724} (even less so at the quantum level, where a confused and scattered literature about gravitational entropy is fashionable, but obscure).   
 
 The discussion above suggests that when macroscopic variables are given, the gradient of entropy itself should single out not just the \emph{direction}, but even the \emph{variable itself} that we call ``time".  This follows from the simple fact that the phenomena that vividly characterize what we call temporal are effects of the entropy gradient. I have repeatedly tried to set up a formalism in which this intuition is realized, but I have not found anything sufficiently clear and convincing to be worth mentioning here.\footnote{Some of my attempts: CR,  General relativistic statistical mechanics, Phys Rev D.87.084055, 2013, arXiv:1209.0065; T Josset, G Chirco, CR, Statistical mechanics of reparametrization invariant systems. Takes Three to Tango, Class.Quant.Grav. 33 (2016) 045005 arXiv:1503.08725;  H Haggard, CR, Death and resurrection of the zeroth principle of thermodynamics, International Journal of Modern Physics D Vol. 22 (2013) 1342007, arXiv:1302.0724; G. Chirco, H. Haggard, A. Riello, CR, Spacetime thermodynamics without hidden degrees of freedom, Phys. Rev. D 90, 044044 (2014) arXiv:1401.5262;  CR, M Smerlak: Thermal time and the Tolman-Ehrenfest effect: temperature as the ‘speed of time’, Class. Quant. Grav 28 (2011) 075007 arXiv:1005.2985; and the paper cited in the previous note (whith the best idea).}  I consider the problem open. 
  
There is only one tool worth mentioning, that allows us to see how a single time with some thermodynamical features could emerge from a fully covariant (hence without any variable that is preferred \emph{mechanically} as a time variable) theory.  Here it is:

\paragraph{Thermal time in relativistic gravity.}   A preferred temporal variable is singled out by the statistical state in which a covariant universe happens to be\footnote{CR, Statistical mechanics of gravity and thermodynamical origin of time, Classical and Quantum Gravity, 10, 1549 (1993); ``The statistical state of the universe", Classical and Quantum  Gravity, 10, pg 1567 (1993).}.  The hypothesis that his is related to temporal phenomenology is called the ``thermal time" hypothesis.  It is essentially the idea that since the ``upper level" aspects of temporality are grounded in the uncertainty of the statistical description of the world underpinning thermodynamics, it is the statistical state of the universe that singles out what we naturally denote a clock time.  The basic idea is simple. In a non-relativistic hamiltonian system, the time evolution is generated by the Hamiltonian $H$ and Gibbs equilibrium states are distribution on phase space in the form $\rho=exp\{-\beta H\}$.  This logic can be reversed. If for some reason we know that the system can be described (possibly relatively to us) by a  distribution $\rho$ on the phase space, then this \emph{defines} (up to scale) a flow on phase space. It is the flow generated by $H_{Therm}=-\frac1\beta \ln\rho$. The parameter of this flow is called thermal time.  The thermal time of a Gibbs state is proportional to the non-relativistic time.  In the general case, where no preferred time variable is given, it  \emph{defines} a time flow.  In collaboration with the mathematician Alain Connes\footnote{A Connes, CR,Von Neumann algebra automorphisms and time versus thermodynamics relation in general covariant quantum theories, Classical and Quantum Gravity, 11, 2899 (1994).}, we have extended this definition  to the general covariant quantum domain.  In this case, the flow is the Tomita flow of the state on the non-commutative observables algebra.  Here it is quantum non-commutativity that underpins the indetermination at the root of (the higher level aspects) of temporality.  Remarkably, a theorem by Connes shows that this flow is unique up to inner automorphisms of the algebra. I view these constructions as intriguing hypothetical suggestions, not yet cleanly developed. 

\pausa 

\paragraph{Is the time arrow perspectival?} 
Finally, I'd like to mention a highly speculative idea that is rooted in the recognition of how perspectival is our world view.  I am not sure this idea is correct (I see counter arguments), but it might be, and I am a bit in love with it.  The full phenomenology of the orientation of time is rooted \emph{uniquely} in the existence of a ubiquitous and consistent entropy gradient in the universe we have access to.  From the point of view of mechanics (classical or quantum, relativistic or quantum gravitational), this appears to be just a contingent fact about our universe.   Among all the innumerable possible solutions of the equations of motion, we happen to be in one with very marked and uniform entropy gradient, at least around and apparently everywhere in the vast portion of the universe we have access to. 

This can be taken as just a fact of the world.  But good science is often born from asking question about how come some facts are what they are.   Not always can we answer, but often we do find enlightening answers.  If we assume that in some distant past the universe was in a low entropy macroscopic state (this is sometimes denoted the ``past hypothesis"), and otherwise quite generic in terms of the equilibrium distribution determined by its dynamics, then we get a good account of our universe.  But this sounds like a partial answer at least because it relies on two assumptions quite in tension with one another: the genericity of the micro-state within the micro-state versus the radical non genericity of the macro-state. 

There is an interesting hypothesis that has not yet been sufficiently explored, that could shed light on this question.  Consider the question of what is the definition of entropy that we use when we observe that entropy grounds the orientation of time.\footnote{The entropy $S(m)$ can be defined as the logarithm of the number of micro-states that are in the same macro-state $m$, namely that have given values of the macro-variables. But what \emph{is} a marco-variable?   Equivalently: heat and work are two forms of energy, what distinguishes one from the other? 
There are different possible definitions.  Each leads to more or less useful results.   One possibility is anthropocentric. We can call macro-variables those we have access to, and can act upon.   This leads to constructing thermodynamics as a theory about what \emph{we} can do to, and get from, a physical system, given our (limited) means of interaction with it.   This is a useful instrumentalist framework.  The resulting notion of entropy depends on ourselves. David Albert used to ask: does  my teacup cool down because of what I can do to it, or the information I have about it?   The rhetorical question is slightly misleading, because the teacup behaves in the way it behaves irrespective of us, but if we ask questions about special variables, then the answers depends on the variables we have formulated the question about.   Yet,  there is a  subtle weakness in such operational approaches.  They are based on assumptions about the agent and the effect of its actions. The agent can record measurements, intervene on the system and the system reacts \emph{after} the intervention, and so on.   These assumptions may appear to be harmless,  but they are not.  For instance, if we are interested in how can a direction of time emerge in thermodynamics from an underpinning time reversal classical dynamics (we definitely are!), we cannot pre-answer the question by assuming that the effect on an intervention follows it in time.   This is what we want to account for, and we cannot account for it if we presuppose it. 
A second possibility is to allow \emph{any} arbitrary choice of macro-variables. This is viable, but of little interest for anybody incapable of accessing or acting on those variables.  
A third option is to identify as macro-variables a set of variables with special dynamical properties. One possibility is to take only constants of motion in the (unperturbed) dynamics of the system.   Another is to pinpoint sets of variables that have an autonomous dynamics, namely whose dynamical behaviour has aspects that can be predicted particularly well, even with small knowledge of the rest.    
Finally, a fourth possible definition of macro-variables gets rid of the crude anthropocentric aspects of the first, but retains its relational aspects.  When two systems, say $A$ and $B$ interact, they generally do so via a subset of their variables.   We can call these  the macro-variables, relative to $B$.  This choice defines the thermodynamics of a system \emph{relative} to a second system.} Commonly, this is based on considering as macro-variables the collection of variables that we commonly use to describe the world, which are those we access: an insignificantly small subset of the degrees of freedom of the universe.  We can get rid of anthropocentrism by realizing that entropy is relative to {\em any} physical system $O$: macroscopic variables relative to $O$ are those $O$ interact with.  This observation immediately opens a possibility: that what is non-generic in the macro-state of our universe is not the dynamical micro-state of the universe (this \emph{is} generic), but rather the subsystem $O$ with respect to which the entropy is being considered. The entropy gradient, that is, could be ---like the rotation of the sky--- real but perspectival.\\[1cm]

\pausa

I close this Seminar on Time by summarizing a few points reached about the nature of time: 
\begin{itemize}
\item In the presence of relativistic (classical or quantum) gravity, the basic mechanical aspects of nature do not single out any preferred time variable. Dynamical relations can be expressed as relations between physical quantities. 
\item No unitarity is required for a quantum theory of gravity. Unitarity is required if there is a global time translation symmetry. The quantum theory is  well defined and internally consistent also in the absence of this symmetry. 
\item Even in the absence of a preferred time variable, (classical and quantum) relativistic physics describes processes, happenings, events, change.   To conceive these as ``static" or ``a block" is meaningless.  Presentism is incompatible with modern physics but this does not imply a naive ``block universe" version of  Eternalism.  It rather requires us to have more flexibility with our intuitions and conceptual structure. 
\item Time-oriented phenomena concern macroscopic variables and the macroscopic description of a system.  They are all sourced by the entropy gradient. The existence of a persistent and ubiquitous entropy gradient in our universe is the source of all phenomena that define what we call the flow of time.  The entropy gradient is a property of the particular contingent solution in which our universe finds itself. The universe was strongly out of equilibrium in the past (in spite of the fact that matter alone was in equilibrium), due to the initial smallness of the scale factor. It persists because typical thermalization times are very long, compared to our scale. 

\item Traces, memories, causation, measurements, clocks, are all phenomena that require dissipation and are sourced and oriented by this global entropy gradient. 

\item The openness of the future is a real perspectival phenomenon that characterizes the perspective of deliberating agent as we are. 

\item When we say that something causes an effect, we are implicitly using an interventionist logic: we compare what can happen with and without the intervention. In doing so, we  break time reversal invariance by keeping the past fixed. We do so because doing otherwise would violate the Second Law. Hence it is thermodynamic irreversibility that underwrites the fact that effects follow causes in time (rather than preceding them in time). 

\item The reason the future is open for us and for which we are interested in causal patterns is because we ourselves are deliberating agents, time oriented by the entropy that underpins our very existence. 

\end{itemize}

\pausa

This Seminar has brought to light the role of the deliberating agent ---rooted in the dissipative processes that underpin our mental processes--- in contributing to the vivid sense of the openness of the future, and the ``passage" of time.  

Kant considered time to be an a-priori condition for intuition. In a naturalized version of his observation, we can say that the contingent existence of a thermodynamical arrow of time in our  environment is a necessary condition for us to know: because knowledge is a natural phenomenon in natural beings like us, and knowledge requires senses, memories, thoughts, and so on, which are all macroscopic phenomena that involve dissipation, which in turn can only happen in a local entropy gradient.

This is the reason why it is so difficult (but not impossible) for us to conceive the non-oriented nature of time: because our very thinking is the child of the orientation of time.  It is one of the products of the  disequilibrium.  We always make the mistake of thinking of ourselves as different from the world around us, looking at it from the outside. And we constantly mistake our perspective for an absolute. 

These observations point to the theme of the next Seminar:  the perspective of the subject. 

\vspace{1cm}

\section{Perspective} \label{subject}

\epigraph
{\em  
\noindent  
Zhuangzi and Hui Shi were walking on the bridge over the River Hao. Zhuangzi observed: 

--- Those little fishes that wander around calmly and without haste are happy fish!\\ 
To which, Hui Shi objected: 

--- You are not a fish; how do you know how a fish is doing? \\
Zhuangzi replied: 

--- You are not me, how do you know what I know about how a fish is?\\ 
 But Hui Shi insisted: 
 
 --- I'm not you, obviously I don't know about you; but you are not a fish and that is enough to deduce that you don't know what a fish likes. \\ 
 But Zhuangzi concluded: 
 
 --- Let's go back to where we started. When you said ``how do you know how a fish is doing?", you knew I knew. I knew it, here, above the River Hao.}
 {Zhuangzi, 17.}

 \noindent   
The last Seminar ended pointing out the relevance of the perspective of the deliberating agent for understanding the openness of the future.  The general observation is that knowledge is always embedded in the natural world, because the holder of knowledge is itself part of nature.  Therefore knowledge is perspectival: it is a form of information that a part of nature has about the rest. It is itself a relation.  

Here I present a few simple consideration that physics can offer, on how knowledge itself can be understood as an integral part of nature, and the subject that holds knowledge can be considered as a physical system like any other, just with its own peculiarities.  I start with three simple observations that can represent steps towards the (in my opinion not yet unachieved) objective of a fully naturalistic understanding of knowledge: the tricky game of understanding what is understanding.    Then I discuss the issue of the accessibility of subjectivity, taking inspiration from the charming text in the epigraph. 

\paragraph{Information.} ``Information" is a word with a wide range of uses and meanings, sometimes very distant from one another.   These can be roughly organized hierarchically, ranging from simple quantitative measurements of correlations or quantum entanglement, all the way up to semantic, mental, social, or even military uses.  The confusion between different uses generates ambiguities and obscurity.   

In its widest use, which underpins (without exhausting) all others, information can be defined in the way Shannon defined the notion he called ``relative information", or ``mutual information".   Unlike other uses of the term ``information", this is a directly physical notion, that does not involve anything mental or semantic. Yet, it is a physical ingredient commonly underpinning most higher level uses of the term.\footnote{CR, Relative information at the foundation of physics, in \emph{It from Bit or Bit from It? On Physics and Information}, A Aguirre, B Foster and Z Merali eds., 79-86 (Springer 2015). arXiv:1311.0054.} 

It is defined as follows.   Consider two physical systems, say $A$ and $B$, and assume that each of them can be in different states.  Let $N_A$ be the number of states in which $A$ can  be and $N_B$ the number of states in which $B$ can be.  A priori, the couple formed by the two systems can be in $N_A\times N_B$ states.  

Consider a concrete situation in which there is a physical restriction in the number of states the two systems together can be.  For instance, say a ball  $A$ can be in the box 1 or in the box 2 and similarly a ball  $B$ can be in the box 1 or in the box 2.   A priori there are four possibilities where $A$ and $B$ are, respectively, in the boxes $(1,1), (1,2), (2,1)$ and $(2,2)$.  Hence $N_A\times N_B=2\times 2 = 4$. But say there is a chain that ties the two balls which is too short for the balls to be in opposite boxes, so that the configurations $(1,2)$ and $(2,1)$ are physically impossible to realize.  In this case the number $N$ of configurations available to the two balls is only two,  not four.    Hence $N<N_A\times N_B$.  Shannon calls ``relative information" the quantity
\be
       I=\ln(N_A\times N_B)-\ln N,
\ee
which quantifies how much each system is constrained by the other.   If this relative information is not vanishing, then knowledge of the state of one of the two systems `informs' about the state of the other.   In this sense, the information so defined captures a common use of the expression ``information":  If there is the chain and one ball is in one box, then for sure the other one is is there as well, hence we have ``information" about the second by just seeing the first.   The first carries ``information" about the second. 

In a probabilistic context, the same idea is expressed by the correlation between variables of the two systems if $a$ is a variable of one system, $b$ a variable of the other, and $p(a,b)$ the probability of having the couple of values $(a,b)$, then the mutual information depends on the correlation
\be
C(a,b)=p(a,b)-p(a)p(b),
\ee
where $p(a)=\sum_b p(a,b)$ and $p(b)=\sum_a p(a,b)$ are the marginals.  

Defined in this manner, information is nothing else than a measure of the correlation between systems.  But notice that it is nevertheless a subtle notion, because it is modal.  It does not refer to what is actual. It refers to what is possible and what is not possible, in the sense of what is allowed by physical constraints and what is not.   It is not a primitive notion, since to be defined in this way it requires a rich framework: notions of systems, states, physical possibilities, physical constraints. 

\paragraph{Meaning.} The second term I discuss is ``meaning".  This again is a loaded term used in a variety of contexts with a variety of (indeed) meanings, that have at most, a ``family resemblance" among them: from the meaning of an expression, to the meaning of life, from the meaning of a term in an equation, to the meaning of a word in a foreign language. 
In many of its uses, the word ``meaning"  appears to be profoundly remote from the physical account the world, as to appear to belong to a different realm, radically incommensurable with the physical reality.   So much so, that ``meaning" is something indicated as a mark of what necessarily separates us from the physical world.\footnote{Recently Chomsky wrote a paper arguing that Large Language Models do not understand the meaning of the text they produce.   If our best understanding of the meaning of a linguistic expression is, \emph{\`a la Wittgenstein}, its use, then this is certainly not the case!} 

There is a simple bridge that can dispel this sense of separateness.\footnote{CR , Meaning and Intentionality = Information + Evolution, in Wandering Towards a Goal, A. Aguirre, B. Foster and Z. Merali editors, Springer 2018.  arXiv:1611.02420.} The bridge is provided by the combination between the physical notion of information described above and the basic insight of evolutionary biology.  

According to this, once certain structures have formed, the effect of natural selection is to favor the abundance of those that have certain traits.   These traits are thus those that favor abundance.  This overall process justifies a language where we call ``useful" these traits.  Since natural selection of structures is compatible with a physicalist account of the world, this shows that final causation, so understood, is also compatible with a physicalist account of the world, which by itself knows no a priori finality.  Notice that this is again re-conceptualization:  final causes are not denied; they are just dethroned from fundamentality and understood as effective descriptions of a rich process.  

Let us now combine this Darwinian\footnote{It is actually older than Darwin, as Darwin recognizes in the Origin of the Species. It can be traced to Empedocles, and is discussed in Aristotle's biology.} insight and the physical notion of relative information mentioned above.   A biological organism is ---like any other physical system--- heavily correlated with its outside.  Its internal parts are also heavily correlated among themselves, and biological organisms are heavily correlated among themselves across space and time.  This is simply because of the actual concrete physical constraints and relations imposed by physical laws and contingent facts.   In this vast ocean of correlations, however, there are some that are specifically ``useful" is the sense above, and in fact ``used" by the organisms' biological processes.     

Take a simple example. A single-cell organism has receptors on its membrane that react to the external abundance of a nutrient, triggering the organism's locomotion towards food.  Here, at some point of the process there is a correlation between the internal state of the cell and the distribution of the nutrient around it. According to the definition above, the organism has information about the location of the nutrient.  But there is more than just information: this is useful information, that plays a key role in the cell's aim-oriented behavior.  Information that plays this biological role can be reasonably called ``meaningful" information, in a sense of ``meaning" that is consistent with some common usage of the world. Formally, this can be expressed by saying that the if we replace the joint probability expressing the mutual information by the product of the marginals, the Darwinian advantage decreases.

When we read on our smartphone that there is a good restaurant in the vicinity, and decide to drive there, we are for sure involved in a far more complex process than the bacterium swimming to sugar.  But there are common aspects that cannot be disregarded.    First, the relation between the location of the sugar and the internal state of the cell on the one hand and the location of the restaurant and the internal state of our brain is in both cases an example of physical correlation.  Second, ``aims" as the bacterium drive to eat and our drive to eat are in both cases consequences (by close or by far) of a long evolutionary process that has singled out the corresponding structures and processes,  at the cellular, as well as the neural, cognitive and cultural levels.  In other words, the sugar is meaningful for the bacterium and the restaurant is meaningful for us for reasons that are different but share a common underpinning, and this underpinning shows that there is no chasm in the  reality we know about: just complexity.

\paragraph{Modality from agency and deliberation.}  The definitions of information and meaning given above are purely physical, and underpin higher level uses of these notions.   These observations are consistent with the idea that between physical and mental concepts there is continuity; the second can be built upon the first.   Yet, both information and meanings have a modal aspect. They refer to possibilities, to `how things could be', not simply as they are.  Where does the relevance of this modality come from, in an actual single actual physical world? 

I suspect that the answer is given by the discussion on causality and the openness of the future given in the previous Section.  Darwinian evolution gives rise to agency, namely to behavior of the organisms that   `purposefully' (in the sense discussed above) affects the environment and (later) to deliberating agent whose behavior is determined by explicitly computing the different outcomes of different actions.  As argued, from the perspective of these deliberating agents, the future is open, different possible futures are reachable, hence modality is an essential tool to apprehend the world.  This is not the reason for which causal patterns exist, but it is the reason for which they are relevant from the perspective of the deliberating agent.   Within this perspective, information, meaning, deliberation and agency are fundamental. 

This is true from \emph{within} this perspective.  {\em But,} we \emph{are} within this perspective.   Because our knowledge is embedded.   And this is because there is no meaning in non embedded knowledge: there is no space where it could inhabit. 

\pausa

\subsubsection*{The happiness of the fish}  

Perspectives can differ. But this does not mean that they can't communicate.  In fact, precisely since they are physical, they are accessible to one another.   The inaccessibility of another's mind is a recurring talking point presented against any form of physicalism.\footnote{See for instance Nagel, Thomas,  ``What Is It Like to Be a Bat?". The Philosophical Review. 83 (1974) 435-450; D Chalmers, Facing up to the problem of consciousness. Journal of Consciousness Studies 2 (1995) 200-19.}.   I think that this inaccessibility is a myth.  Minds are no less nor no more inaccessible than everything else in nature.  

 In the delightful dialog in the epigraph of this Seminar,\footnote{Here is the original :
 \begin{CJK}{UTF8}{bsmi}
 . 庄子与惠子游于濠梁之上(1)。庄子曰：“儵鱼出游从容(2)，是鱼之乐也。”惠子曰：“子非鱼，安知鱼之乐？”庄子曰：“子非我，安知我不知鱼之乐？”惠子曰：“我非子，固不知子矣；子固非鱼也，子之不知鱼之乐，全矣！”庄子曰：“请循其本。子曰‘汝安知鱼乐’云者，既已知吾知之而问我。我知之濠上也。”
% 莊子與惠子遊於濠梁之上。莊子曰：「儵魚出游從容，是魚樂也 。」 惠子曰：「子非魚，安知魚之樂？」莊子曰：「子非我，安知我不知魚之樂？」 惠子曰：「我非子，固不知子矣；子固非魚也，子之不知魚之樂，全矣。」莊子曰：「請循其本。子曰『女安知魚樂』云者，既已知吾知之而問我，我知之濠上也。」
And a different translation: 
 Zhuangzi and Huizi were strolling (you 遊) on the dam of the Hao River. Zhuangzi said, “How these minnows jump out of the water and play about (you 游) at their ease (cong rong 從容)! This is fish being happy (le 樂)! ”
Huizi said: “You, sir, are not a fish, how (an 安) do you know (zhi 知) what the happiness of fish is?”
Zhuangzi replied: “You, sir, are not me, how (an 安) do you know (zhi 知) that I do not know (bu zhi 不知) what the happiness of fish is?”
Huizi said: “I am not you, sir, so I inherently don’t know you; but you, sir, are inherently no fish, and that you don’t know (bu zhi 不知) what the happiness of fish is, is [now] fully [established].”
Zhuangzi replied: “Let’s return to the roots [of this conversation]. By asking “how (an 安) do you know (zhi 知) the happiness of fish,” you already knew (zhi 知) that I know (zhi 知) it, and yet you asked me; I know (zhi 知) it by standing overlooking the Hao River.”
(Zhuangxi, 17. Translation in Meyer, “Truth Claim”, 335, modified as in Lea Cantor (2020) Zhuangzi on ‘happy fish’ and the limits of human knowledge, British Journal for the History of Philosophy, 28:2, 216-230.)   
\end{CJK}
}   taken from the great old Chinese philosophy text, Hui Shi objects to the apparently innocent observation by Zhuangzi that the fish are happy: happiness is a {\em subjective} matter and as such it is hidden from the outside; only the fish can know if they are happy, not Zhuangzi, who is not a fish.  This is the same issue as in the famous ``What Is It Like to Be a Bat?" by Thomas Nagel, mentioned in the last footnote and used to claim a pretended insolubility of the mind/body problem within physicalism, on the basis that subjective experience is hidden.  

The first move of the extraordinary reply by Zhuangzi (``You are not me, how do you know what I know about how a fish is?") equates the relation between Zhuangzi and the fish to the relation between Zhuangzi and Hui Shi.   The difference is in degrees, not in nature, because for sure there are differences between Zhuangzi and a fish, but there are also similarities, and for sure there are similarities between Zhuangzi and Hui Shi, but there are also differences.  Hence if Hui Shi considers the fish a priori inscrutable, he must consider Zhuangzi inscrutable as well. 

Hui Shi is forced to concede that his objection implies that he does not know about Zhuangzi. And here comes the devastating second step of Zhuangzi's judo-like move: ``you knew I knew", because this is implied by the very objection you raised!  It had to be so, because otherwise how could the very communication between Zhuangzi and Hui Shi have happened in the first place, without Hui Shi assuming to know something that was in the mind of Zhuangzi? The very fact that the two are talking, namely communicating, shows that communication between subjectivities is possible. If communication is possible, there is no {\em a priori}  obstacle for Zhuangzi to know something about the fish's happiness.   

Of course Zhuangzi may have the fish happiness wrong. Or Hui Shi might have arguments showing that the fish are actually distressed, or incapable of happiness by the nature of their species, but there is no \emph{a priori} barrier in talking about the happiness of the fish. Hui Shi (and Nagel) are wrong in pretending there is an a priori barrier, precisely as there is no barrier in Zhuangzi and Hui Shi understanding each other and knowing about each other.

These observations do not imply that Zhuangzi knows (or may know) \emph{everything} going on in the fish brain. He does not.  But this does not single out subjectivity  from the other phenomena of nature,  nor do we know \emph{everything} going on, say, inside a stone. In a stone, there are vast oceans of complicated molecular and subatomic processes that happen at every moment, and we  have de facto no access to them.  If we have no full access to what goes on in a stone, of course we have no full access to what happens in a brain either. In fact, even introspection does not give us access to what goes on in own own individual brain, either.  There is no a priori difference between a mind, which is the name we assign to what goes on in a brain, and a stone, as far as accessibility to our understanding is concerned.   Of both, we have a permanently incomplete conceptual grasp ---built up from information, knowledge, observation, measurement, communication, images, memories, empathy, emotions and so on--- that connect them with processes in \emph{our own} brain.  This connections inform us about the phenomena that we call physical as well as about the phenomena we call mental. As Bertrand Russell put it: ``The raw material out of which the world is built up is not of two sorts, one matter and the other mind; it is simply arranged in different patterns by its inter-relations: some arrangements may be called mental, while others may be called physical."\footnote{B Russell, The Analysis of the Mind (Mc Millan 1921).}

Nor do these observations imply that Zhuangzi can have \emph{certainty} about the happiness of the fish.   But we never have certainty about anything else, either.  We have credible and reliable indices, and this is what we call knowledge.   The amount of knowledge we can have about a fish, a bat, including their ``subjectivity", whatever this means, is incomplete and uncertain, but can  be reliable, in the same general manner in which our knowledge of anything else is incomplete and never entirely certain, but can be reliable.  

Nor, for that matter, are subjective experiences more manifest and self-evident to ourselves than any other knowledge we have. I see red.  Are you sure? No actually, now that you say so,  I was wrong: what I saw was rather a violet...    Even the Cartesian evidence of existing (``I am'') is not an immediate ascertainment of a self-evident truth: as Descartes himself puts it, it is the conclusion of a complex intellectual argument that starts from the cultural impulse to doubt.\footnote{The claim of the {\em Discourse on the Method} is not ``I am as this is evident", but rather ``I doubt, but I could not doub without thinking and could not think without being".} Similarly, Heidegger's answer to his `question of being', via the experience of the being that poses this question, resolves the issue into an highly intellectual game, whose pretended immediacy follows, doesn't precede, the full education of a XX century German philosopher.  Subjectivity, in brief, is not hidden  in any more special manner than any other feature of reality.   

\pausa

 \paragraph{The Easy Problem is Hard Enough.}  Adding the ghost of a peculiar a priori inaccessible reality to what is nothing else than the way we denote the complexity of the phenomena happening in the brain, is the kind of mistake that Wittgenstein  warns us against in the Philosophical Investigations: adding entities where there is none, mistaking criteria for symptoms.\footnote{PI, I, 354.}   The happiness of the fish is not a mysterious unphysical reality whose physical manifestations are symptoms. Rather, those physical manifestations are the criteria under which we define \emph{what we mean} by the happiness of the fish.  These physical manifestations can range from the fish calm wandering, to the dopamine in their brain, to particular structures or processes in their (and our) brains, all the way to  dogs wavering tails, or Thomas Nagels telling of his happiness.  The actual connection between these natural phenomena form the (evolving) empirical ground for our concept of happiness.  There is no ``happiness" over and above all this, of which these would be, as Wittgenstein put it, ``symptoms".  

David Chalmers famously distinguished a hypothetical ``hard problem" of consciousness from the (very hard) ``easy problem" of understanding what happens in our brain.\footnote{D Chalmers, Facing up to the problem of consciousness. Journal of Consciousness Studies 2: 200-19, 1995.} The ``hard problem" seems to me to be a false problem stemming from posing the \emph{assumption} of a dualism between a physical reality that can be known, perceived, felt, and a transcendental subject that knows, perceives, feels. This forgets the fact that the subject is itself part of nature, and  what we call its knowledge \emph{can} itself be seen as a physical phenomenon: a physical relation between two physical systems.  The source of the confusion seems to me the initial step: to treat knowledge as intrinsically unphysical.  It is \emph{this} mistake (which I believe reappears even if the most naturalism-oriented of the philosophers, such as Quine) that prevents the pacific closure of the circle between our epistemology and our ontology.  

The point I am making is not that the subjective perspective can be reduced to an objective understanding.  It is the opposite: that any account is perspectival, because there is no knowledge that is not embedded in nature. Therefore any pretended objectivity in knowledge is ultimately equally subjective as any subjectivity.  The same is true for any forms of subjectivity. What is like to be a be a fish? It is like to be whoever is posing this question, up to matters of degrees, and these matters of degrees can be studied.
\pausa

The observation that the subject of knowledge, the owner of a consciousness, the experiencer of senses, impressions, or qualia, is itself part of the natural world ---a world that somebody else can observe--- is certainly far from being new. It runs through the history of common sense as well as the philosophies of several civilizations, as the Zhuangzi testifies.   Yet, the temptation to single out the subject as very special and treat it abstractly ---for one reason or another---  also returns continuously in the history of common sense and the history of thinking.  It is always lurking, and it is a permanent source, I believe, of false problems, artificial implicit dualism and conceptual confusion. 

In the Philosophical Investigations, Wittgenstein writes ``If the formation of concepts can be explained by facts of nature, should we not be interested, not in grammar, but rather in that in nature which is the basis of grammar?—Our interest certainly includes the correspondence between concepts and very general facts of nature. [...] But our interest does not fall back upon these possible causes of the formation of concepts; we are not doing natural science; nor yet natural history". As for myself, instead, I \emph{am} doing natural science, and after I learn (from him) to chase the ghost out of the black box, the scientist in me is curious of how the black box works.  

We use words to refer to processes and relations, not magical entities.  But words themselves are processes and relations, not magical entities.  Physics can contribute to dispel some of the apparent magic, by revealing connections and links between what we call the physical world and the world of meaning, language, norms, knowledge, information, understanding and similar.

\newpage

\section{Without certainty and without foundation}  \label{pp}

\noindent  It is time to bring a few threads together. What follow are a few reflections on  results and tools of my trade, presented without any pretension.\\[.3cm]

\subsubsection*{Concepts evolve} 
%\begin{figure}[h]
\vspace{.5cm}
\begin{flushright}
\includegraphics[width=4.87cm]{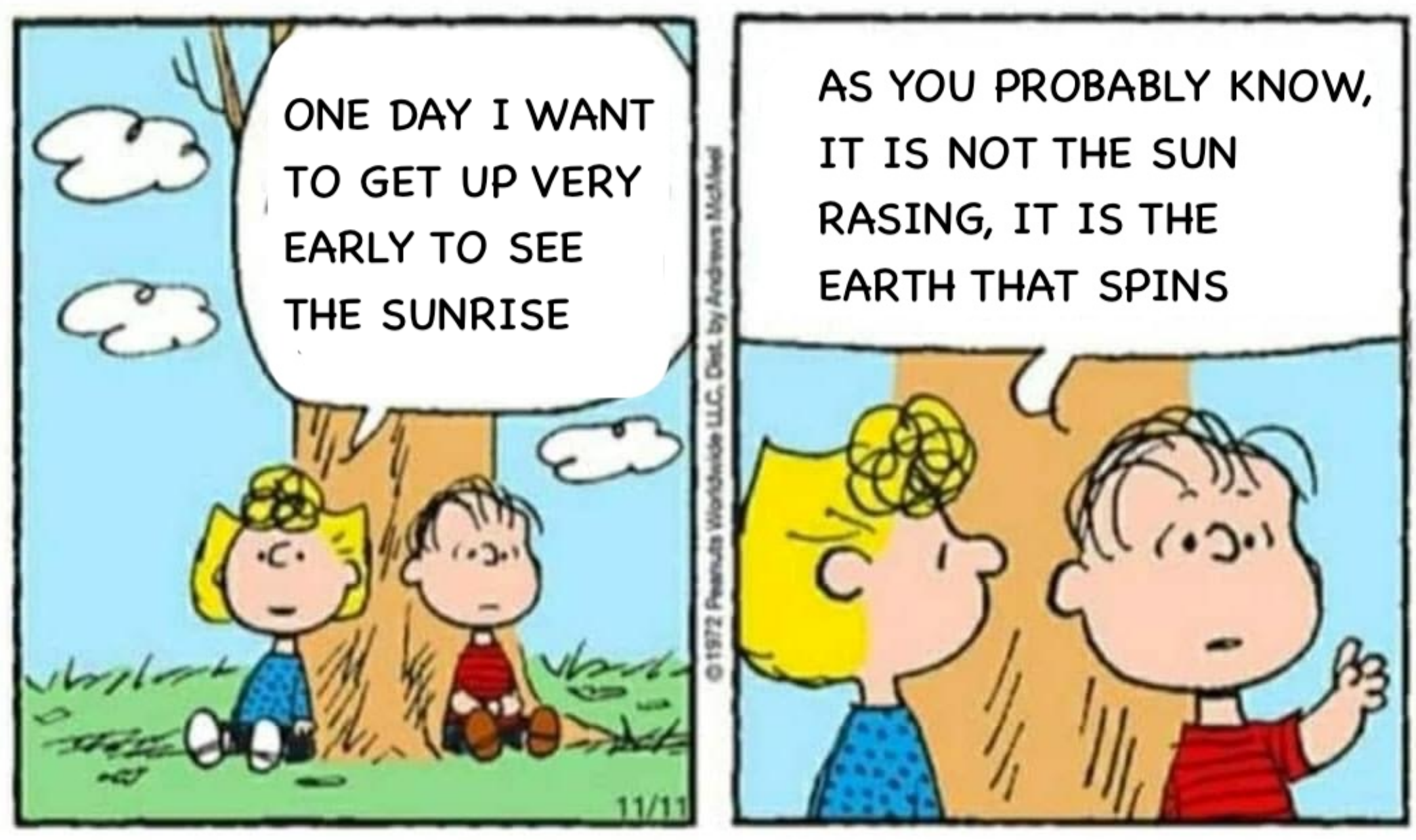}
\includegraphics[width=5cm]{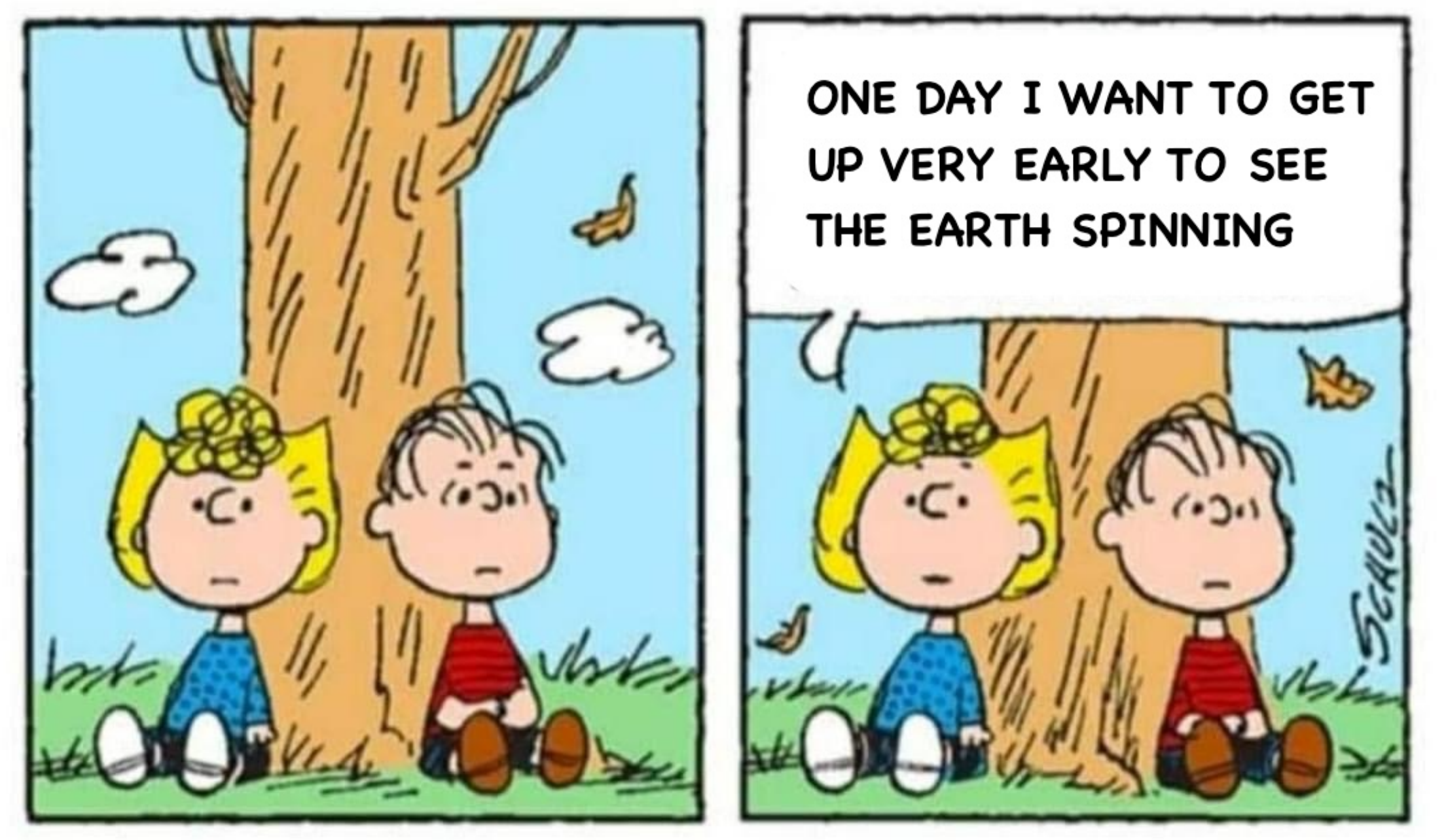}
\end{flushright}
%\end{figure}

\vspace{.5cm}

A recurrent theme in these Seminars has been that the meaning of concepts changes when we learn something new.   

When we all got convinced by Copernicus that a sunrise is not the Sun actually rising or doing anything else, but is rather a perspectival phenomenon due to the Earth rotation, we did not conclude ---in anguish--- that ``there are no sunrises".  Or, worse, that a sunrise ``is an illusion".   Rather, we simply accepted that we could better understand the phenomenon that we denote ``sunrise".\footnote{The point is developed and applied to a variety of situations in \href{http://philsci-archive.pitt.edu/18837}{CR, The Old Fisherman's Mistake}, Metaphilosophy 53 (2022) 567-746.}  We re--conceptualized ``sunrise": we changed its definition---not anymore the Sun actually raising.  Nothing of this re-conceptualization is in the way of any function that sunrise plays in our lives or any development of a science of sunrise.

I find that a common obstacle to understanding comes from hanging on to a specific meaning of concepts. If we insist that by ``sunrise" we \emph{mean} a Sun actually moving upward, by ``free will" we actually \emph{mean} that we can govern Nature from outside it, and by ``time" we actually \emph{mean} a universal one directional ``flow",  by ``state of a system" we mean the ensemble of properties that a system has by itself, we throw ourselves into conceptual corners that prevent us from better understanding nature.  This is the biggest obstacle that I have found in developing a quantum theory of gravity.  Sometimes, careful conceptual analysis is like wallowing in the mire of our wrong prejudices. 

When we better understand the stratification of time, its rich layered structure, or when we trace the phenomenology of its orientation to the entropy gradient (outside and inside ourselves), we are not deeming the flow of time or the openness of the future as illusory.  We better understand what they are.  

In slowly deciphering the biochemistry and the thermodynamics of life, we do not show that life is illusory or non existent: we understand how it works.   What we find to be illusory is only the previous belief of a fundamental dichotomy between living and not living matter.   Similarly, in slowly deciphering the functioning of our brain, we do not show that subjective phenomena (or `consciousness', which is the modern word, pretending to be scientific, for `soul') are illusory: we are simply beginning to understand how they work.  To anguish for a feared loss of free will is like anguishing for the loss of sunrises after Copernicus. 

In the example of the fish in the previous Seminar, I argued that there is no ``happiness" of the fish over and above the criteria for calling them happy; these are not, as Wittgenstein put it, ``symptoms" of something else ineffable.    This does not mean that there are no genuine happiness, consciousness, interiority...  It only means that we can better understand these phenomena by not viewing them as  unphysical ghosts. Physicalism is not in the way of the richness of our interiority.  It provides a way to ground our understanding of it.     

Concepts came after, not before, knowledge.  Howard Stein puts it like this: ``Reason, or the ratio of all we have already known, is not the same that it shall be when we know more"\footnote{H. Stein, How Does Physics Bear upon Metaphysics; and Why Did Plato Hold that Philosophy Cannot be Written Down?, Colloquium Talk at the University of Chicago.  See A.W. Carus, The Pragmatics of Scientific Knowledge: Howard Stein's Reshaping of Logical Empiricism, The Monist, Vol. 93, No. 4, Philosophical History of Science (2010), 618-639.}.  Conceptual clarity can be misleading, when the concepts we have are non-suitable for the new domains we would like to extend them to.   When we ask a question like what is the ``true meaning" of a concept we use (time, space, evolution, causality, free will, physical observable, particle, perception,  data, universe, knowledge, self, energy, or whatever else...) we are asking the wrong question. These are vague expressions whose use is variable and drifts.  The most fertile concepts in physics (just think about `Energy' and `Entropy') have been constantly redefined and still maintain an aura of vagueness.   A good part of science has been, and still is, a constant questioning and readjusting of its own conceptual foundations.  Quantum gravity, the central topic of my scientific work, requires this drift and we are better equipped to work on it if we are aware of this fluidity. 

Science at its best challenges our conceptual structure and forces us to modify it, to find one better adapted to phenomena.  This is the science of Copernicus, Kepler, Newton, Darwin, Lavoisier,  Einstein and so many others.  The scientific revolution of the XX century has been a feast on conceptual subversion.  The feast is not over.\\[1cm]

\subsubsection*{Reality as interaction} 

\epigraph{\em 
`I say therefore that what by nature can act on another or suffer even the slightest action from another, however insignificant it is, and even if it happens only once, this alone can be truly real.  I therefore propose this definition of being: that it is nothing if not action (\textgreek{δύναμις})'.}{Plato. The Sophist.}

Reality is definitely more subtle than the simplistic materialism of particles in space, or fields in space.  What seems to me to emerge from the developments in fundamental physics of the last century is a pervasive relational core.   

GR leads us to a relational understanding of spatiality and temporality.    The Newtonian space and time  structures can be seen as aspects of an entity --the gravitational field-- but this is a dynamical entity on the same footing as all others, while spatial and temporal localizations only make sense relationally. The ``where" and the ``when" of any event are not properties of the event by itself: they \emph{only} refer to  how the event is placed with respect to other events.    

This understanding of space and time is close to the Aristotelian one\footnote{Time is  a counting of events (\textgreek{Φυσικά} IV.1-14), and, as already mentioned, the location of an object is the innermost  boundary of the surroundings (\textgreek{Φυσικά} IV.5). Both are relational notions.}, but has a radical side, especially as far as time is concerned: it contradicts our intuition of an absolute (non-relational) universal ``now".   If by ``time" we insist in denoting this intuition (as Mc Taggart does\footnote{John McTaggart, The Unreality of Time, Mind 17 (1908) 457--474.}), then time in this sense does not exist (as Mc Taggart concludes).  But to do so is like claiming that sunrise does not exist because the Sun doesn't move.  Sunrises exist: they are just a more complicated (relational!) phenomenon than what we used to think them to be before Copernicus.  Temporality is real: but it is a more complicated relational phenomenon than what a naive intuition may suggest. 

But the strongest invitation to think in relational terms comes from quantum phenomena.  Niels Bohr's insight on quantum phenomena is that they can only be understood by including the measuring apparatus.  If we clean this key insight from its unnecessary anthropocentric and instrumentalist overtones, we catch what seems to me the central discovery of QM:  to understand a quantum phenomenon we need to refer it to the phenomena it affects.  The contingent property of a system \emph{are} the ways the system affects other systems.  In other words: the way a system affects another system is not a symptom about its state; it is what we mean when we talk about its state, or an indication on what else to expect.\footnote{M Dorato,  Bohr meets Rovelli: a dispositionalist account of the quantum state. Quantum Studies: Mathematics and Foundations, 7 (2020).233--245.}

It is  thanks to this shared common relational aspects that GR and QM are consistent: quantum gravity, as I understand it, is built on the identification of  the quantum and the general relativistic relationism: identifying Heisenberg cuts with spacetime boundaries opens the way to the conceptual structure discussed in the Seminar \ref{qs}.   The quantities that describe the world (when we do not neglect its relativistic gravitational features nor its quantum properties) capture how a spacetime process affects another spacetime process. 

\pausa

The relevance of relations is not a novelty in science.  Relativization of physical quantities previously regarded as absolute is a leading thread through the history of physics.\footnote{The daily rotation of the sky is not a feature of the sky: it is due to our relation with it: we look at it from a spinning planet.  Galileo's major insight, which underpins classical mechanics, is that velocity is relational: there is no meaning in the velocity of a single body, unless it is explicitly or implicitly defined relative to some other body. Colours are not properties of an object: they are complex phenomena emerging form the interaction between the object, light and the peculiar receptors in our eyes. Voltage at a point of an electric system is not an absolute quantity: is relative to another point.   Gauge theories are written in terms of connections, which define how internal spaces are oriented relative to one another.   Charge is defined by its action on other charges. Entropy, as discussed above, is best understood as relational.   And so on, the list could be very long.}  Not is the relevance of relations a novelty in philosophy\footnote{There are major relational aspects in the worldviews of Leibniz, Kant, Nietzsche, Whitehead, Nagarjuna, and many others. Surprising relational views can be found even where one least expects them, as in the epigraph of this section, where Plato gives a completely relational definition of being.}, ancient thinking\footnote{Think for instance at the mutual-dependence doctrine in the \href{https://archive.org/details/Brihadaranyaka.Upanishad.Shankara.Bhashya.by.Swami.Madhavananda/mode/2up}{Second Chapter of the  Brihadaranyaka Upanishad},  (``The earth is like honey to all beings and all being are like honey to the earth"), 378--390.}, or theology\footnote{For example: G Maspero, The Cappadocian Reshaping of Metaphysics, Relational Being (Cambridge University Press 2024). Ch. Yannaras, Relational Ontology (Holy Cross Orthodox Press, 2011.) }.  In other sciences and other domains, relational thinking is even more ubiquitous.\footnote{Chemical properties of elements and substances are virtually always describing how elements and substances interact with others. Most features of a species in biology cannot be understood without considering  the other species it interacts with.  In economy, an individual agent would make no sense unless we thought it interacting with the others.  Psychology is largely concerned with the relations between individuals, and so on. Music is not just sound: it is produced by the internal relations between frequencies and rhythms and by the relations between these and the cultural memories and expectations in our brain (On this: CR, La musica; Dialogo con Erik Battaglia, in CR, Lo sapevo qui sopra il fiume Hao (Solferino 2023) 29-35).}
Yet, the relational character of fundamental physics appears to have more radical consequences than this ubiquitous relationism, because it undermines the hope that  physics could provide a substratum on which other relations could be built, the bricks forming relational structures.  If one thinks that the possession of variable intrinsic properties is essential to the identity of an entity and no entity can exist if it does not have an intrinsic identity, then  the relativization extends to the identity of physical systems and quantum processes.\footnote{M Dorato, Bohr meets Rovelli: a dispositionalist account of the quantum limits of knowledge, op. cit..}  We describe a quantum system (or process) $A$ in terms of the way it affects another system $B$.  But of course $B$ is also a quantum system, hence we understand its properties in terms of the way these affect a further system, and so on.  

Given a physical system $O$, the rest of the universe can be described in terms of the ways it affects $O$, or ``from the perspective" of $O$, but we cannot describe the full universe $U$, intended as the collection of everything, because by definition there is no system outside $U$, with respect to which there are properties of $U$.\footnote{This does not regard the scientific field that goes under the name of ``quantum cosmology", because this research --despite its name-- is only concerned with the large scale degrees of freedom of the cosmos, not with all its degrees of freedom. See F Vidotto, Relational Quantum Cosmology, in The Philosophy of Cosmology (Cambridge University Press 2017) 297--316.}  

The resulting perspectivism runs deep, because of the observation at the end of Section \ref{q}: relativity iterates. That is, there is no contingent statement about the world that escapes it, including relational statements like ``this fact obtains, relative to $O$". This statement is {\em itself} relative\footnote{This has been recently discussed in detail in Timotheus Riedel, Relational Quantum Mechanics, quantum relativism, and the iteration of relativity, Studies in History and Philosophy of Science 104 (2024) 109–118.}  (to $O$, or to other systems).   

Devoid of non-relational contingent properties, entities themselves do not seem to have much of an independent existence. This takes us close to Plato's relational definition of being in the epigraph, or to Nāgārjuna's radical version of dependent origination\footnote{J Garfield (translation). The Fundamental Wisdom of the Middle Way: Nāgārjuna's Mūlamadhyamakakārikā,   (Oxford University Press  1995)}.  

But if we only understand something in terms of something else, don't we remain without a solid starting point?  If the physical entities themselves can only be understood relatively to something else, don't we fall into a groundless circularity? 

I think we do.  And I  think that there is nothing wrong with this.   In fact, it seems to me that it is the other way around: it is the myth of a substance with autonomous reality, independent from anything else, or the idea of the ultimate ground of our worldview, to be the source of our confusion.  In other words, yes, the science described above appears to be foundation-less.  But I do not think that this is a defect. I think it is a virtue.  Before discussing this point, to which I will come back in closure, however, let me shift topic and collect a few considerations on the \emph{methodology} of science, where relationism again plays a role.\\[.1cm]

\pausa
\vspace{.1cm}

\paragraph{Cumulative science.}  The observation of the relevance of conceptual evolution in science is not an argument for discontinuity.  In fact, I find far more continuity in real science than in the way science was presented to me.  The two physics revolutions of the XX century, relativity and quantum, have sounded as a  shocking rupture because the  success of Newtonian physics had nurtured the misguided delusion of a definitive understanding of the ultimate nature of reality.  Finding it factually wrong (Mercury does not move according to Newton's laws) has shifted all the attention to discontinuity, and has ended up nurturing the equally misguided idea that a given scientific theory does not offer a credible world picture.  

What needs to be questioned is only the adjective ``ultimate", not the credibility of the picture.  The Newtonian picture of the world remains of course perfectly viable (all our engineers keep using it) and good.  What has been put into question is only the illusion that it was an ultimate universal account of reality.  Not being universally true does not mean it to be false.  Newton's theory is fabulously informative, predictive and insightful; it is not exact nor complete, but nothing of what we humans do is ever exact or complete anyway.  The Newtonian conceptual tools capture real aspects of the world. If to be like Newton theory is to be ``false" then our entire knowledge is false, as current theories are  likely not in better shape than Newton's theory in this regard.  If this is what we mean by ``false", I genuinely see no reason to care about ``truth".

Relativity and quantum have developed conceptual tools,  laws, ways of thinking about reality, which also work where Newton theory does not. They are not incompatible with Newton's theory where this works.  What we have learned with relativity and quantum, if we do not get trapped by the delusion of \emph{ultimate} knowledge, is not something ``instead": it is just something ``more".  Science is cumulative. 

I have argued that precisely the same is true even for Aristotelian physics\footnote{CR, ``Aristotle's Physics: A Physicist's look" Journal of the American Philosophical Association, 1 (2015) 23-40, arXiv:1312.4057}.  It is very good physics, powerful and correct, in its domain of validity; it stands in relation to Newtonian physics precisely as Newtonian physics stands in relation to relativity: as a good approximation, expressed in appropriate concepts, in the regime for which it was developed\footnote{Motions of bodies of different densities immersed in fluids with higher or lower densities than the bodies, such as water or air, and subject to gravity, friction and buoyancy.}, as well as a step in the development of more far-sighed pictures. Most of past knowledge remains in our baggage, sometimes simply re-expressed in a slightly new language.  

Continuity runs even deeper, historically.   The internal structure of quantum theory parallels classical Hamiltonian mechanics, relativistic mechanics is a minor modification of non relativistic mechanics.   Field theory, a radical departure from particle mechanics was anticipated by the motion of the molecules of the surfaces of a fluid. And so on.  Tools, structures, ideas, objects, theoretical terms, pass though Kuhnian revolutions intact or only slightly modified.  What does so is not just structures --as it has been claimed--:  the Sun of the Egyptians, the one of Ptolemy, Copernicus, Newton, Einstein, the one of the nuclear reactions and the one of the theory of stellar evolution, is always the same Sun. We are simply learning  more about it.   To say that it is a different entity seems to me, once again, to confuse our always imperfect and incomplete knowledge with definitions. 

\paragraph{How do we learn, I. From past knowledge.}

I think that dismissing continuity is currently misleading research.  My colleagues\footnote{There are exception; a particularly interesting one is Aur\'elien Barrau. See A Barrau ``Anomalies cosmiques: La science face \`a l'etrange" (Paris: Dunod 2022).} mostly hold a naive version of Popperian-Kunhian philosophy (the views of scientists, whether they like it or not, are impregnated with philosophy) that nurtures the misleading illusion that new theories should be guessed, fished out of the blue sky, and everything not falsified could be considered equally likely, equally in need of testing.  This is an attitude that dismisses acquired knowledge as unreliable, in exploring new domains. 

I think that history of science teaches us the opposite.  Developments in fundamental physics have never come from guessing. Either they were forced by unexpected data (Kepler's ellipses by Mars anomalies, Maxwell by Faraday's experiments, quantum mechanics largely by atomic spectroscopy, quantum chromodynamics by the hadron's zoology...), or they were built from established knowledge by ``taking it seriously"\footnote{The expression, in this context and in this sense, is by Feynman.}.  

An example of the second case is Special Relativity, forced by Einstein's reliance on Galilean relativity of velocity and on Maxwell equations. Following the Popper-Kuhn `vulgata' popular today among physicists, Einstein should have guessed a modification of Maxwell equations, or given up relativity of velocity in high speed regimes.  Instead, he trusted past scientific knowledge and got rid of the apparent contradiction between Galileo's and Maxwell's insights by maintaining reliance on both and instead modifying the conceptual structure: changing the meaning of simultaneity.  

General relativity as well is a product of the reliance on acquired knowledge: special relativity, Newtonian gravity and using Maxwell theory as a model.\footnote{I discuss this point in CR, General Relativity. The Essentials (Penguin 2021), Available at \url{https://www.cpt.univ-mrs.fr/ rovelli/GR.pdf}.}  Similarly, Copernicus builds virtually entirely on Ptolemy: on facts, style, tools... everything except for a small foundational conceptual detail.\footnote{The two books, the Almagest and the De Revolutionibus are extraordinarily similar in all their aspects.}

In all these and similar cases, advances happened largely without new data.\footnote{Copernicus had virtually no more data than anybody else in astronomy in the previous millennium.  I believe Einstein's claim that the Michelson-Morley result was not important for him, because the Michelson-Morley result is indeed \emph{not} important for deriving special relativity. Mercury's perihelium anomaly worked as a validation, certainly not as a source, for General Relativity. Newton's logical arguments build mostly on Galileo and Kepler (he definitely used new data, but, again, as a validation).} 
They did not emerge from magical intuitions, but from a careful scrutiny and critical evaluation of internal apparent contradictions, incongruences, or strange coincidences in the available knowledge.\footnote{For Copernicus, the strange fact that in Ptolemy the deferent of some planets and the epicycles of the others have the same period. For Newton the incongruence between Galileo's and Kepler's laws, for special relativity the inconsistency of the Maxwell equations with the established relativity of velocity, for general relativity the inconsistency of the Newton force with special relativity.}  They show that  major ``revolutions" in the history of fundamental physics are far more conservative than the way they are advertised.  They did not stem from discarding previous knowledge: on the very contrary, they built on previous knowledge.\footnote{Copernicus found the obvious (a posteriori, of course) solution to the puzzle raised by the Ptolemaic coincidences.  Newton, the obvious (a posteriori, of course) way to combine Kepler and Galileo. Einstein, the  only reasonable way to have both: Maxwell equations and relativity of velocity, and so on.  The fabulous case of electromagnetism is in between.  Faraday's (and previous) data were essential, but the final synthesis by Maxwell was only constructed on completing for coherence previously discovered laws.}   New physics stems from previous knowledge, from taking it seriously, as reliable understanding of reality.

\pausa

Why was then so hard to take these steps? 
Because each of them implied \emph{also} a radical rejection of something that was not previously under question: the very definition of what is the task of astronomy for Copernicus, the culturally established distinction between Heavens and Earth for Newton, simultaneity, and then familiar geometry for Einstein...  My examples come  from physics and astronomy, which are familiar to me. But I trust this to be valid  widely in science.  Darwin is of course a major example of drastic (in good part data driven) equally \emph{conceptual} revolution in biology. 

Progress has not come about within a clear methodological and conceptual framework by discarding past theories in defect of verification for better ones pulled out of the blue sky.   It is entirely the other way around:   progress has come about \emph{changing} methodological and conceptual framework, \emph{relying} on the scientific validity of core insights of past theories, that were {\em not} in defect of verification, and replacing them with new theories that are certainly not pulled out of the blue sky, but rather \emph{motivated} as solutions of puzzles raised by acquired knowledge, reached at the price of conceptual changes. 

This is an account of scientific progress different from the one I learned at school.  I think that a certain sterility of recent fundamental physics is due to being misled by badly digested philosophy of science: discarding past knowledge as unreliable and trying to fish randomly ``new" theories out of the blue sky, without questioning the current conceptual structure. Thousands of Physical Review D papers do just that, and lead nowhere.   Specifically, Kuhn's emphasis on discontinuity has misled many into devaluing the relevance of past knowledge for making sense of what we do not yet know.  Popper's emphasis on falsifiability, originally just a demarcation criterion, has been flatly misinterpreted as an evaluation criterion.  The combination of the two has nurtured a methodological confusion: the idea that all unproven ideas are equally plausible, all unmeasured effects are equally likely to occur.\footnote{While theorists and experimentalists interested in fundamental physics have spent the last decades focusing on searching something beyond General Relativity and the Standard Model, Nature has constantly disappointed them: ironically, the recent momentous results in experimental basic physics are all rebuttals of today's freely speculative theoretical physics. Consider gravitational waves, the Higgs, the absence of supersymmetry at LHC, and violation of the Bell inequalities. All these are confirmations of old physics and disconfirmations of widespread speculation. In all  cases, Nature is telling us: do not speculate so freely. The detection of gravitational waves, rewarded by the last Nobel Prize in fundamental physics, has been a radical confirmation of century-old general relativity. But it is more than this. The recent nearly simultaneous detection of gravitational and electromagnetic signals from the merging of two neutron stars (the event called GW170817) has improved our knowledge of the ratio between the speeds of propagation of gravity and electromagnetism by something like 14 orders of magnitudes in a single stroke. This has ruled out a great many theories put forward as alternatives to general relativity, studied by a large community of theoreticians over the last few decades. The well publicized detection of the Higgs particle at CERN has confirmed the Standard Model of particle physics, against scores of later alternatives that have long been receiving much attention. The absence of supersymmetric particles has disappointed a generation of theoretical physicists expecting to find them. The verification of entanglement even at large distances, and many other quantum phenomena, always presented as surprising, are just confirmation of what I studied at school 40 years ago.  Perhaps, Nature's repeated snub of the current methodology in theoretical physics should encourage a certain humility in our philosophical attitude.}

The work of a theoretician has been misinterpreted as consisting in exploring arbitrary possibilities, as anything that has not yet been falsified might be equally plausible. This is the current `Why not?' ideology: any new idea deserves to be studied, just because it has not yet been falsified and may be a Kuhnian discontinuity, unpredictablable on the basis of past knowledge; any experiment is equally interesting, provided it tests something as yet untested.  

The quantum gravity theory illustrated in Seminar \ref{qs} stems from the opposite logic: quantum gravity does not require ``new ideas": it requires thinking carefully about what we have learned about the world in deciphering relativistic gravity and quantum phenomena, and asking which conceptual rearrangements are required to make them consistent.\footnote{In other words, I think that our understanding of how science works has moved \emph{too} far from Newton's ``whatever is not deduced from the phenomena must be called a hypothesis; and hypotheses, \emph{whether metaphysical or physical}, or based on occult qualities, or mechanical, have no place in experimental philosophy. In this philosophy particular propositions are inferred from the phenomena, and afterwards rendered general by induction." Newton makes it too simple here, of course, but we have moved far too far, seems to me, and this may be a reason of certain sterility and a remarkable series of failures of contemporary fundamental physics.}

\paragraph{How do we learn, II. Analogies and metaphors.}

Nothing of the above implies that progress is easy, of course.  There is no methodology for discovery.  (Where there was, progress has happened already.)   
It seems to me that the way we learn is the blind search for a delicate balance. The balance between what to bring with us and what to leave behind. What we bring with us is what guides us. Einstein used Maxwell's equations, and Galilean relativity. Copernicus used Ptolemy's book. These are the maps, the rules, the generalities we trust because they have worked well.

Yet, we know that we must leave something behind. Anaximander leaves behind the idea that all things fall parallel. Einstein leaves behind the idea that all clocks tick in unison... If we leave behind too many things, we don't have the tools to move forward, but if we bring too much, we won't find the gaps toward understanding... I don't think there are recipes, except for trial and error. Trying and trying again. That's what we do. The long study and the great love.

We combine and recombine in different ways what we know, seeking a combination that clarifies something. If they hinder, we leave out pieces that seemed essential before. We take risks, but with caution. We get familiar with the edge of our knowledge and frequent it for a long time, back and forth, groping for gaps. We build new concepts trying new combinations.

We build by analogy.  Our new concepts are taken from old concepts, re-adapting and modifying them metaphorically. Newton's ``forces" are borrowed from the everyday experience of a push. Faraday's electric and magnetic fields, extended in space, are stolen from the farmers.  Einstein understood that time passes sometimes slowly and sometimes quickly, but haven't we known this from experience since ever?

Of course, an electromagnetic field is not a field of wheat, Einstein's time dilation is not due to boredom, there is no one pushing and pulling in the force of gravity; but the analogies are evident. Analogy is taking an aspect of a concept, reusing it in another context, preserving something of its meaning while letting go something else, so that the new combination produces new and effective meanings.\footnote{William Whewell: ``there is a New Element added to the combination, by the very act of thought by which they were combined” (in The Philosophy of the Inductive Sciences, Founded Upon Their History London: John W. Parker. II, 48).   Whewell calls this “colligation”: the mental operation of bringing together a number of empirical facts by “superinducing” upon them a conception which unites the facts and renders them capable of being expressed by a general law.  He claims that a large part of the history of science is ``history of scientific ideas,” that is, the history of their explication and subsequent use as ``colligating" concepts.  See: L.J. Snyder, William Whewell, The Stanford Encyclopedia of Philosophy, E.N. Zalta and U. Nodelman eds.} Scientific thought makes good use of logical and mathematical rigor, but this is only one of the two legs that have brought it to success; the other is the creative freedom allowed by the continuous evolution of its conceptual structure, and this grows through analogies and the inexhaustible power of recombination they offer.\footnote{The West has effectively used the creativity of thought through analogies to construct new concepts in every generation, ultimately bequeathing the magnificence of scientific thought to today's global civilization. However, it is curiously the East that recognized first and with greater clarity that thinking grows more through analogies than through syllogisms. The logic of argumentation based on analogies was already analyzed by the Mohist school and implicit in one of the greatest books of humanity, the Zhuangzi.}  This seems to me how the best science works.\footnote{I believe this is also how the best art works. Science and art involve the continuous reorganization of our conceptual space, of what we call meaning. What happens when we react to a work of art is not in the object itself, even less so in some mysterious spiritual world; it is in the complexity of our brain, in the kaleidoscopic network of analogical relationships through which our neurons weave what we call meaning. We are engaged because it takes us slightly out of our habitual sleepwalking, reigniting the joy of seeing something new in the world. It is the same joy that science brings.  Vermeer's light opens our eyes to a resonance of light that we had not yet been able to grasp; a fragment of Sappho (``sweet-bitter is Eros") reveals to us a world of reconsidering desire; Anish Kapoor's black void disorients us like the black holes of general relativity; like these, it suggests that there are other ways to conceptualize the intangible fabric of reality.}

It is the ability to change the organization of our thoughts that allows us to make the major leaps forward.\footnote{Think about what Copernicus did. Before him, the world was divided into two major categories: earthly things (mountains, people, raindrops...) and celestial things‚ objects like the sun, moon, and stars. Earthly things fall, celestial things revolve in circles. Earthly things are perishable, celestial things eternal. It is so reasonable that it takes courage to propose another way of organizing the world. Copernicus does it. His cosmos is divided differently. The Sun stands alone in its own category. The planets are all in the same category; the Earth is just one of them, with everything it contains, so mountains, people, and raindrops belong in the same category as those dots in the sky called Venus and Mars... the moon, well, it is in a category of its own: everything revolves around the sun, but the moon revolves strangely around the earth.} The path between observing and understanding can therefore be very long.  The reason is we are always trapped in innumerable wrong assumptions, that structure our concepts, and we always make the mistake of taking concepts more seriously than reality.  Science is therefore sometimes the same activity that Wittgenstein described as (his) philosophy:  ``Philosophy is a battle against the incantation of our intellect, by means of our language"\footnote{PH, I, 109.} and: ``do not think, observe!" \footnote{PH, I, 66.  
In PH, II, 12, LW writes: ``If the formation of concepts can be explained by facts of nature, should we not be interested, not in grammar, but rather in that in nature which is the basis of grammar? Our interest certainly includes the correspondence between concepts and very general facts of nature. (Such facts as mostly do not strike us because of their generality.) But our interest does not fall back upon these possible causes of the formation of concepts; we are not doing natural science; nor yet natural history..." Here, on the contrary, I \emph{am} doing natural science, and therefore interested in ``that in nature which is the basis of grammar".}. 

\pausa

%There is no reason why it should continue to work beyond that. Dividing everything into earthly and celestial objects may be suitable for daily life, but not for understanding the cosmos and our place in it.

\paragraph{Models as relations.}

Philosophers tend to take scientific models very seriously.  I think that reality is more complex than our individual models.  Models are tools, not mirrors of reality.   By this I do not mean that reality is ``beyond our reach".  Nor do I mean that we do not have good pictures of reality. I mean the opposite: we have a far better understanding of reality than any one of our models.   We hold a rich and complex understanding of reality, composed of a plurality of layers and it makes no sense reducing it to individual models.  These layers, in turn, are ways we are in relation with processes. 

Take a simplest part of reality you can imagine, say a bottle of water resting on a table.\footnote{I have stolen this example from Erik Curiel. Thanks Erik.}  We can model the water as a litre of liquid with a certain mass and volume. We can do better and view it as a thermodynamical system characterized also by temperature and pressure.   But of course we know that there is rich hydrodynamics gong on, micro-turbulence, and more.   There are peculiar phenomena at the interaction between the water and the bottle.  There is a complicated relation between the surface, the air, the vapour.   There are the billions of molecules bouncing around.  There are living organisms swimming in the water. These have a complicated biology, of which we know a lot.   There is the nuclear physics of the nuclei of the atoms.   Not to mention the fact that the bottle is there because we drink, so our understanding of it, including the reason for which it is on the table is based on this function.  And of course the color of the label reflects the specific culture of the country where it was produced; and the fact that it is a commercial bottle of Italian source water on a US table is indicative of a certain organization of the international capitalistic consumer market,  of the US commercial and political relations with Italy, and so on and so on. 

All this is part of our knowledge and understanding of the bottle.  Dismissing a part of this as less relevant is nonsensical.  Some aspects are more fundamental than others with respect to certain given concerns. From the perspective of fundamental physics, we do not get much reduction either, however, because the water is a family of excitations of quantum fields, in a theory (quantum field theory) of which we have confused conceptual  understanding (because of the confusion around quantum theory) and which we know is badly incomplete (because of quantum gravity). 

Each of the layers and each perspective from which we have partial understanding of the bottle has its own level of coherence and utility.  Each of them represents one possible way we and other parts of reality can interact with the bottle.  So, in this sense, each regards a way the bottle interacts, manifest itself, including, as I have emphasized, its most elementary physical description. 

Yet, nobody has any  doubt that we are talking about the same bottle, and nobody should, because it {\em is} always the same bottle, like the teacher in the class and the kind partner at home are the same persons.  There is nothing strange in viewing the same object, person, or phenomenon from multiple perspectives. We do so regularly in life. The multiplicity does not break coherence, it only adds richness. But all this does undermine the idea of a complete and ultimate description of anything.  

And yet, the different perspectives on the bottle are coherent.   They do not contradict one another. If they appear to, this raises our curiosity and we think that there is something missing and there is something  we can understand better.   Some articulation between levels are foggy for us, while others are notoriously worked us pretty well. An example of the first is the articulation between physical and sociological aspects of the same object, an example of the second is the connection between chemical and physical properties.   But far from testifying to the disunity of science (as some have argued), they only testify to its complexity and richness; their coherence testifies for the unity, not the disunity, of science.   `Mysteries' are juicy pointers to our empirical ignorance and --even more interestingly-- conceptual confusions.

Some perspectives are more fundamental from their own point of view, which is legittimate, as others are equally more fundamental from their different points of view, equally legitimate. Because ``fundamental" is a relative word.   It is true that in a sense the bottle is ``nothing else than atoms", as it is true that the very notion of ``bottle" is perspectival, or social, or else.   I see no contradiction, unless in terms of questions about ultimate foundations and ``ultimate reality", which sound meaningless to me. 

Furthermore, physics is modal. It lists possibilities, that can be used to derive statements of the form ``if A, then B".  Models are therefore statements about conditionals, possibilities.  They are meant to be applied repeatedly and to capture regularities, not individual pictures of phenomena. 

A good model is a tool for seeing a system or a phenomenon from a perspective; it captures some accessible aspects of it and allows us to think about what goes on in a useful manner.  In particular, (in physics) it provides relations between accessible variables, which allow predictions, extrapolations, and sometimes control.  Since ``accessible" refers to something external (whatever has the access), any model is intrinsically relational.  This is why confusing the model with the reality is a mistake.    The variables of the physical systems are the ways it interacts with other systems. Here we meet again the perspectivism from which I started this Seminar. 

\pausa

\subsubsection*{What Science does}

\epigraph{\em 
I believe that ideas such as absolute certitude, absolute exactness, final truth, etc. are figments of the imagination which should not be admissible in any field of science. [...] This loosening of thinking (Lockerung des Denkens) seems to me to be the greatest blessing which modern science has given to us. For the belief in a single truth and in being the possessor thereof is the root cause of all evil in the world.}{Max Born. Nobel lecture}

In the course of history, science has constantly redefined itself, along with its goals\footnote{A few examples from physics and astronomy. In light of Hipparchus and Ptolemy's extraordinarily successful predictive theories, the goal of astronomy was identified in finding the right combination of circles that give the motion of the heavenly bodies around the Earth. Contrary to expectations, it turned out that Earth was itself one of the heavenly bodies. After Copernicus, the proper goal appeared to be to find the right combination of moving spheres that would give the motion of the planets around the Sun. Contrary to expectations, it turned out that abstract elliptical trajectories were better than spheres. After Newton, it seemed clear that the aim of physics was to find the forces acting on bodies. Contrary to this program, it turned out that the world could be better described by dynamical fields rather than bodies. After Faraday and Maxwell, it was clear that physics had to find laws of motion in space, as time passes. Contrary to assumptions, it turned out that space and time are themselves dynamical. After Einstein, it became clear that physics must only search for the deterministic laws of Nature. Contrary to expectations, it turned out that at best we can make probabilistic predictions. And so on. Here are some sliding definitions for what scientists have thought science to be: deduction of general laws from observed phenomena, finding out the ultimate constituents of Nature, accounting for regularities in empirical observations, finding provisional conceptual schemes for making sense of the world... and so on.}, methods, and tools\footnote{The historical development that led to modern science is a very slow progress in finding {\em tools} for better understanding and dealing with the world.   Observations, critical thinking, rational debate, use of theoretical terms, materialism, geometry, systematical use of doubt, empiricism, experiments, algebra and advanced math, physical laws, realism, statistics, verification, falsification, operationalism, and so on (in rough historical order of appearance); these are tools of the trade of a scientists (not always compatible among themselves, but this is the nature of tools).   Tools evolve with the trade, and I believe they are still evolving in the scientific practice (not always in a good direction).}. 
This very flexibility has been an ingredient of its success.  Science has never been a project with a methodology written in stone and a settled conceptual structure. It is rather the result of a series of steps in the ever evolving endeavor to make sense of the world, by the concrete social and biological (hence physical) critters we are: to organize our understanding of reality. Science has repeatedly violated its own rules and its declared methodological assumptions.  The methodology, the philosophical perspective, have sometimes followed, not preceded, the growth of knowledge. 

Science is thus a historical phenomenon where we make up and modify the rules ``as we go along", as we do in anything else.\footnote{PI, I, 83.} Not just an evolving body of empirical information we have about the world and a sequence of theories, but rather the evolution of the  notions, the  concepts, we use to organize our thinking. It is an ongoing search for the best conceptual structure for grasping the world, at each given stage of knowledge. The scientific enterprise aims at building conceptual understandings of reality, as reliable as possible, constantly revised on the basis of new information,  data,  theoretical developments, cultural exchanges, and else.    Our conceptual framework is neither definitive nor the only possibility: it is only what evolution has led us to patch together in order to navigate our everyday affairs. This is why science repeatedly changes the order of things. 

\paragraph{Knowledge as imperfect relation.}

The idea of a disembodied rational agent  (whether single scientist or community of thinkers) collecting data and organizing them in theories, neglects two facts. First, the rationality of the agent is itself in evolution:  the modification of the conceptual structure is  achieved from within the agent's thinking, as a sailor who has to rebuild his own boat while sailing, in the celebrated simile by Neurath often quoted by Quine.\footnote{W V O Quine, Word and Object (Cambridge, Mass.: MIT Press 2015).}   Second, and more importantly, the subject that develops science and holds knowledge is not a disembodied rational entity: it is itself a \emph{physical} system, part of the world, obeying the same laws it is trying to unravel.  Hence knowledge itself must be understandable --not only in how it forms, but also in what it is-- as an aspect of the world picture being developed.  I am not sure we have clarity about this, yet.  The small steps described in Seminar \ref{subject} are certainly insufficient.  Consistency requires the circle between the subject and the object of knowledge to genuinely close. Our current lack of a clear understanding of what ``to understand" actually means, renders the entire project confused.

The knowledge science produces is a historically located phenomenon regarding peculiar correlation between peculiar physical systems --socially organized mammal brains-- and various aspects of the rest of reality these have easy access to.  Internally, it is structured by a conceptual scaffolding formed by notions that cannot be a priori: they have evolved with our brain, our society and our culture, and evolve with the evolution and interchanges of knowledge itself.  Externally, it must be a peculiar form of correlations between our brains, our books, and the rest of the world. 

Knowledge, in other words, is a relation that we should be able to grasp in natural terms.  The information contained in our knowledge is a complex form of meaningful (in the biological sense discussed in Seminar \ref{subject}) correlation between physical systems.  A family of correlations between the content of the memory in our brains or the texts in our books, and facts of nature.  We  close the circle by recognizing that what is established in understanding is a form of these same correlations that we find between physical systems.  

This observation seems to me to demolish the intuition that our subjectivity is peculiar and unaccountable from a physicalist perspective. Not because it is easy to understand how our mental processes work. It is not easy, it is hard, because they are complex phenomena. But because the subjective perspective is not extraneous to physicalism. In fact, it is the only option available to physicalism, once relational nature of reality and the consequent perspectivalism are taken into account.   There is no third person view of reality.  There are only first person views.  This is not panpsychism: psychic phenomena require the full complexity of our brain.  It is rather the realization that we are in front the world not as an ``I" in front of a ``it", but rather as an ``I" in front of a ``you".   The rest of the world is like us: a physical system.  Where each part interacts with another as we do interact with others.  We should address the world as we address our brothers and sisters. Aware that there are plenty of aspects of it, as of them, that we do not reach. 

The world we observe, study, and get emotionally entangled with, is the reflection of other systems upon us: a part of the networks of correlations that weave reality. When we interact with the world we always interact with partial aspects of it.  We can see the forest, or the trees, or the leaves of the trees, or its biochemistry, or its atomic physics, or its quantum gravitational granular spatiotemporal structure, or --in another direction--, we can see the economical role of the forest, the lovers walking it it, the spinning planet on which it sits, the vast galaxy it belongs to, lost in the  ocean of other galaxies.   We organize all these and other partial views of the world according to diverse interesting criteria of fundamentality.   I do not see any claim of one single entry point to be self standing to hold.  The entry point is always an unexplained blind spot in the picture.   The coherence between levels we keep finding, the lack of contradiction, the absence of hard dualities, all this does not change the fact that our understanding remains fragmentary and any pretension of deriving all from one entry point leaves out most of what we know.   

This is what seems to me we actually mean by ``reality", in our normal legitimate use of the word.  It is  what we describe and interact with, including the awareness that we have a partial grasp of it, as we deduce from commonly seeing something that somebody else does.  

\pausa

\subsubsection*{Foundations} It seems to me that the radical perspectivism that we find at the core of physics, the realization of the continuously drifting aspects of our own basic conceptual schemes, and the realization of the physical nature of our very knowledge, converge in inviting us to be  suspicious of any idea of ultimate substance, ultimate description, preferred foundation, solid ground for knowledge, possibility to access ultimate truths, certainties, or secure methodology for knowledge.  Any model we have leaves out more than what it captures.  All theories we have are incomplete.  All attempts to establish a solid ground have ended up showing loopholes and flaws.  Whatever we say, feels as if leaving out something unsayable. Notions such as ``foundational", ``ultimate" are themself profoundly perspectival: dependent on which questions one prioritizes.  Foundations are useful only if they are plural.  It seems to me that the physical, historical, contextual and accidental aspects of our knowledge indicate that it is wiser for us to renounce ambitions of a primary philosophical foundation, whether this is on data, phenomena, reality, perceptions, laws, substance, a fundamental ontology, introspection,  relations\footnote{I find the most illuminating insight in Nāgārjuna's Mūlamadhyamakakārikā to be the observation that mutual dependence is itself mutually dependent (``emptiness is empty"). It allows a radical anti-foundationalism, which I find extreemely liberating, both intellectually and ethically--ultimate questions have no meaning.}, or whatever.   

The reverence to the notion of truth on the one hand, and Popper's falsifiability on the other, have shadowed the key aspect of knowledge, emphasized by Ramsey and (for its dynamical side) by De Finetti: reliability comes in degrees. It can be extremely high, and as such sufficient for us.  This makes certainty superfluous.  I see certainty as both superfluous and unreachable: a delusional mythology.  There is no reason to keep searching for what cannot be found and is of no use.

In life, we never require certainty: a high degree of reliability is enough.  It suffices to enter buildings, convict criminals,  rely on a flight schedule,  trust our wealth to a bank, build a bridge and walk over it, love the people we love..., even if bad surprises and judicial errors happen, flight schedules contain mistakes, bridges fall, and love betrays.  

Skeptics periodically remind us that there is nothing certain in any knowledge.  Whatever we think we know might  turn out to be mistaken\footnote{My father discovered as a young adult that his name was not the one he thought it was.}, illusory, a product of a confused memory or a cultural fabrication.  No amount of attention, rigor, collective collaboration, a priori thinking, can screen us with absolute certainty from errors and illusions.  You might just be dreaming of reading this text, or having proved a theorem: you actually may be a dreaming butterfly as in the famous apologue of the Zhuangzi.\footnote{Once, Zhuang Zhou dreamed he was a butterfly, a butterfly flitting and fluttering about, happy with himself and doing as he pleased. He didn't know that he was Zhuang Zhou. Suddenly he woke up and there he was, solid and unmistakable Zhuang Zhou. But he didn't know if he was Zhuang Zhou who had dreamt he was a butterfly, or a butterfly dreaming that he was Zhuang Zhou. Zhuangzi, II.}  This takes nothing away from our efforts in wisely using attention, rigor, collective collaboration, analytical thinking, efforts to make up some precise (mathematically rigorous) versions of imprecise concepts. But these are only imperfect tools, to be handled with wisdom, aware of their limits and dangers: they give us no guarantees.  Our knowledge is valuable \emph{because} reliability has degrees: reliability can be high and valuable, also in the lack of certainty.   The geographical map we are using may contain errors and is for sure incomplete, but we do not gain anything by throwing it away. There is no reason to worry of being a dreaming butterfly.   

We better rely serenely on structurally uncertain knowledge.  It is the best we can do, and it is actually very good for us.  Tribunals, flights, bridges and love are valuable, notwithstanding uncertainty. So is our uncertain knowledge.  We live in this space between ignorance and  certainty.  The two extremes are of no interest. What matters to us is the space in between.    

This affects any attempt of a foundation.   I find anything related with Certainty, Truth, Reality as it really is, Foundation, search for an Ontology, the Best System, and similar, of little interest, if these worlds are capitalized, namely not taken in their weak deflated imprecise common use.  In their capitalized version, they have no meaning I can grasp.  They look to me as projections of a psychological or cultural anguish, pathological extrapolations due to the fears of this mammal with an overgrown brain.  

As far as I can see, reality is more tenuous than the clear-cut one imagined by the old physics models; it is made up of happenings, discontinuous events, without permanence, located with respect to one another and  only existing relatively to one another.  It is a fine texture, intricate and fragile as Venetian lace... Its events are punctiform, discontinuous, probabilistic, relative. Our knowledge and our awareness of them is just one among the relations that weave it. 

Yet, of course, the evolving, uncertain, incomplete, weakly founded, circular family of relations that we  put together and call knowledge, is nevertheless a formidable tool at our disposal for  understanding,  navigating and contemplating the network of relations that we call reality. Even if it is, or perhaps because it is, as Japanese Wabi-Sabi, ``imperfect, impermanent, and incomplete".\\[.4cm]

\raggedleft{Verona, London Ontario, Santa Fe, Venice, June 2023--July 2024}

\end{document}